\pdfoutput=1

\documentclass[11pt]{article}

\usepackage[final]{acl}

\usepackage{times}
\usepackage{latexsym}
\usepackage{multirow}
\usepackage{soul}

\usepackage[T1]{fontenc}

\usepackage[utf8]{inputenc}

\usepackage{microtype}

\usepackage{inconsolata}

\usepackage{graphicx}
\usepackage{xcolor,soul}
\usepackage{booktabs}       
\usepackage{xcolor,colortbl}
\usepackage{dashrule}
\usepackage{xspace}
\usepackage{amsmath}
\usepackage{mdframed}
\usepackage[hang,flushmargin]{footmisc}
\usepackage{enumitem}
\usepackage{pifont}
\usepackage[normalem]{ulem} 

\usepackage{arydshln}
\setlist[itemize]{noitemsep, topsep=0pt}
\setlist[enumerate]{noitemsep, topsep=0pt}
\definecolor{paleaqua}{rgb}{0.74, 0.83, 0.9}
\definecolor{forestgreen}{HTML}{006600}

\newcommand{\customsize}[1]{\textsc{#1}}
\newcommand{\msmarco}{\customsize{MS MARCO}\xspace}
\newcommand{\hotpotqa}{\customsize{HotpotQA}\xspace}
\newcommand{\fever}{\customsize{FEVER}\xspace}
\newcommand{\fiqa}{\customsize{FiQA}\xspace}
\newcommand{\arguana}{\customsize{ArguAna}\xspace}
\newcommand{\nq}{\customsize{NQ}\xspace}
\newcommand{\scidocsrr}{\customsize{ScidocsRR}\xspace}
\newcommand{\gptmini}{GPT-4o-mini\xspace}
\newcommand{\gpt}{GPT-4o\xspace}
\newcommand{\ab}{\customsize{AIR-Bench}\xspace}
\newcommand{\beir}{\customsize{BEIR}\xspace}
\newcommand{\mytab}{\hspace*{1em}}

\newcommand{\hlgreen}[1]{\sethlcolor{green!30}\hl{#1}}
\newcommand{\hlorange}[1]{\sethlcolor{orange!30}\hl{#1}}
\newcommand{\hlred}[1]{\sethlcolor{red!15}\hl{#1}}
\definecolor{mygray}{gray}{0.1}
\newcommand{\cmark}{\color{mygray}\ding{51}}%
\newcommand{\xmark}{\color{mygray}\ding{55}}%

\newcommand{\huggingface}{\raisebox{-1.5pt}{\includegraphics[height=1em]{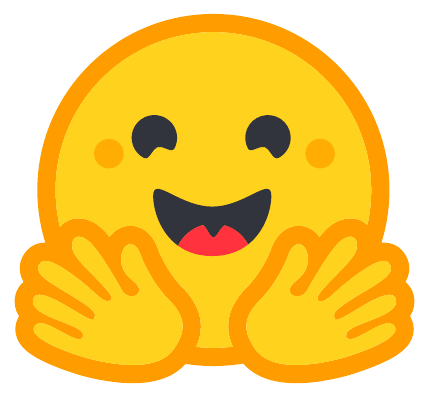}}}
\newcommand{\github}{\raisebox{-1.5pt}{\includegraphics[height=1em]{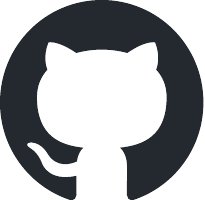}}}

\title{Hard Negatives, Hard Lessons: Revisiting Training Data Quality \\for Robust Information Retrieval with LLMs}

\author{Nandan Thakur\thanks{Equal contribution.} \ \ \ Crystina Zhang\footnotemark[1] \ \ \ Xueguang Ma \ \ \ Jimmy Lin \\[1ex]
David R.\ Cheriton School of Computer Science,\\
 University of Waterloo, Canada \\[1ex]
 \hspace*{-0.8cm}\github{} \textbf{Code:} \texttt{ \url{https://github.com/castorini/rlhn}} \\
  \hspace*{-0.8cm}\huggingface{} \textbf{Dataset:}\texttt{ \url{https://huggingface.co/rlhn}} \\
}

\begin{document}
\maketitle
\begin{abstract}
Training robust retrieval and reranker models typically relies on large-scale retrieval datasets; for example, the BGE collection contains 1.6 million query-passage pairs sourced from various data sources.
However, we find that certain datasets can negatively impact model effectiveness ---
pruning 8 out of 15 datasets from the BGE collection, reduces the training set size by 2.35$\times$, surprisingly increases nDCG@10 on \beir by 1.0 point.
This motivates a deeper examination of training data quality, with a particular focus on ``false negatives'', where relevant passages are incorrectly labeled as irrelevant.
We utilize LLMs as a simple, cost-effective approach to \emph{identify} and \emph{relabel} false negatives in training datasets.
Experimental results show that relabeling false negatives as true positives improves both E5 (base) and Qwen2.5-7B retrieval models by 0.7--1.4 points on \beir and by 1.7--1.8 points at nDCG@10 on zero-shot \ab\ evaluation.
Similar gains are observed for rerankers fine-tuned on the relabeled data, such as Qwen2.5-3B on \beir.
The reliability of LLMs to identify false negatives is supported by human annotation results. 
Our training dataset and code are publicly available.

\end{abstract}

\section{Introduction}

Modern-day retrievers and rerankers are data-hungry, relying on large and high-quality training datasets to accurately retrieve or rerank on challenging domains~\cite{thakur:2021, muennighoff-etal-2023-mteb, chen-etal-2025-air, su2025bright}. 
A typical training dataset for information retrieval (IR) has multiple instances consisting a training query, labeled positive passages, and a set of mined \emph{hard negative passages}. 
Sampling hard negatives has been consistently used in retrieval models to improve downstream retrieval accuracy~\cite[\emph{inter alia}]{karpukhin-etal-2020-dense, xiong2021approximate, qu-etal-2021-rocketqa, moreira:2024}.

More recently, state-of-the-art (SoTA) retrieval models are observed to fine-tune on enormous or large training datasets~\cite{qwen3:2025}. 
While the general notion is that more training data is better, in accordance with scaling laws~\cite{chen:2024,li:2024c,muennighoff2025generative}, we show the contrary:\ fine-tuning on a select few datasets is rather crucial. For example, removing ELI5 surprisingly improves nDCG@10 on 7 out of 14 of the \beir datasets~\cite{thakur:2021} and the average nDCG@10 by 0.6 points.
A similar observation is also made on other training datasets:\
by pruning 8 out of the 15 datasets in the BGE training collection~\cite{li:2024c},\footnote{The pruned dataset contains only 42.5\% training pairs of the original dataset, making it 2.35$\times$ smaller in size.} the E5 (base) retrieval model improves by 1.0 point nDCG@10 on \beir (as shown later in \autoref{fig:beir-leave-one-out-scores}).

\begin{figure*}
    \centering

    \vspace{-1em}

    \includegraphics[width=\linewidth,clip,trim=15 0 15 0]{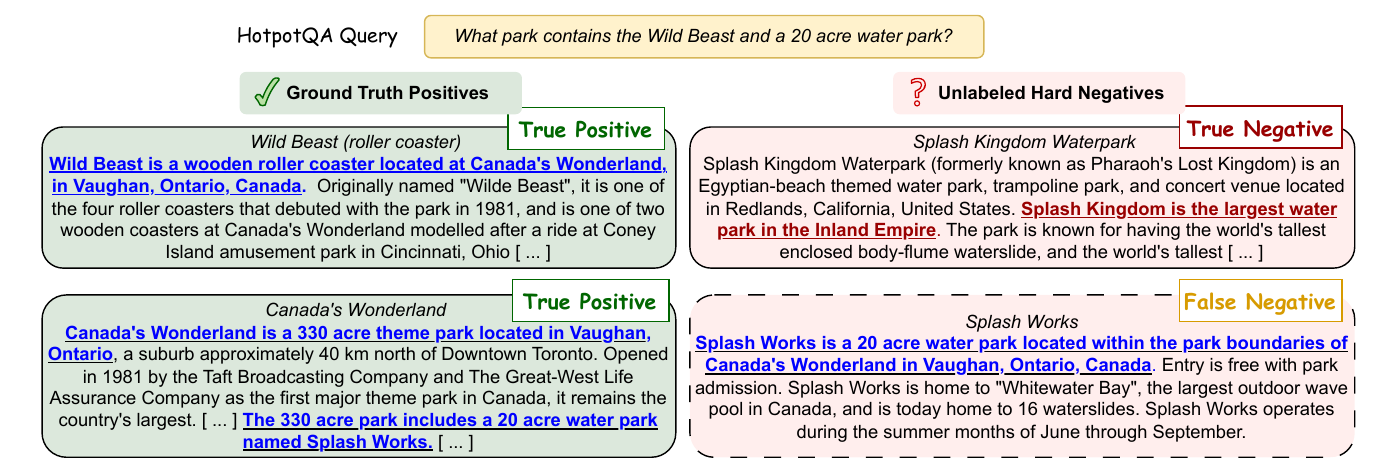}

    \vspace{-1em}

    \caption{Example of a training instance (query, ground truth positives, and unlabeled hard negatives) with detected false negatives taken from \hotpotqa. The false negative passage (\textit{Splash Works}) is \textbf{mislabeled} as it is relevant in answering the user's query. 
    The relevant parts of the text useful in answering the query are highlighted in \textcolor{blue}{\uline{\bf blue}}.
    }
    \label{fig:enter-label}
\end{figure*}

The above observation reveals a non-negligible amount of ``false'' or mislabeled data is mixed in the current training datasets,
that not only adds unnecessary training cost but also hurts the model training.
\emph{How can the ``false'' data be eliminated?}
We approach the issue from the perspective of \emph{false negatives} (example in \autoref{fig:enter-label}), 
specifically, by proposing \textbf{RLHN} (\underline{R}e\underline{L}abeling \underline{H}ard \underline{N}egatives) utilizing a cost-effective framework with large language model (LLM) cascading~\cite{chen:2023} to accurately identify and relabel false negatives (at a data sample level).
We choose to look at false negatives since it is a systematic pitfall from how retrieval training data is constructed:\footnote{In contrast to ``false positives'', that only results from mistakes of human annotators in training datasets.}~As long as there are \emph{unjudged} documents used as negative examples,
the issue of false negative persists,
which is especially severe for big sparsely-annotated datasets, such as \msmarco~\cite{nguyen:2016} or \nq~\cite{kwiatkowski:2019}.

The issue of false negatives has been noticed for long --- \citet{qu-etal-2021-rocketqa} distill knowledge from a cross-encoder to alleviate their impact. \citet{moreira:2024} filter potential false negatives based on relevance score to the query. However, the former solution does not curate or clean the training datasets and is based on the assumption that the cross-encoder is more robust to false negatives than retrieval models. 
As we will show in Section \ref{sec:results}, albeit smaller, inferior training data also negatively affect cross-encoders. The latter solution is based on the assumption that the relevance scores of false negatives are systematically higher than 95\% of the positive scores, which does not consider score variance at the level of a data instance.

\begin{figure*}[t]
\centering
\begin{center}
    \includegraphics[trim=20 10 20 10,clip,width=\textwidth]{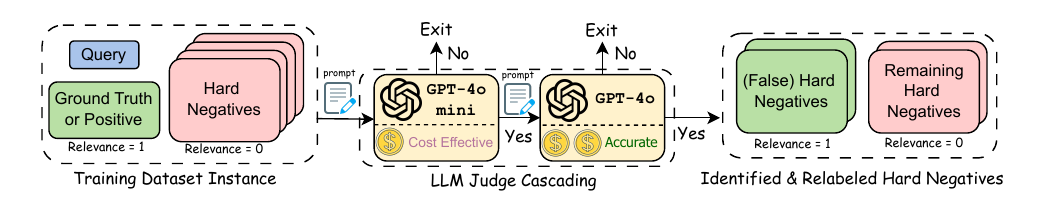}
    \caption{Flowchart for \textbf{RLHN} (\textbf{R}e\textbf{L}abeling \textbf{H}ard \textbf{N}egatives): (1) Provide the query, ground-truth or positive passages, and hard negative passages from a training instance as input, (2) Prompt a cost-effective LLM judge (e.g., \gptmini) and evaluate whether any hard negative is misclassified, (3) If yes, repeat the prompt with an accurate LLM judge (e.g., \gpt) (4) Output the relabeled hard negative passages (which are found relevant) and either remove them or relabel them as ground-truth passages in our experiments.}
    \label{fig:diagram}
\end{center}
\vspace*{-\baselineskip}
\end{figure*}

We use an LLM cascading framework to alleviate ``false negatives''. The first stage employs \gptmini, a cost-effective LLM, to identify false negatives in all training instances. Next, the detected instances with false negatives are relabeled with a more reliable judge, \gpt.
We observe a maximum of 56\% of training pairs in \msmarco can contain false negative documents, to a minimum of about 3\% in \scidocsrr. The framework is better illustrated in \autoref{fig:diagram}.
With the false negatives detected, we compared three data modification approaches:\ (i) \emph{remove}: discarding the whole training instance, (ii) \emph{remove~HN}: removing only the false hard negatives, and (iii) \emph{relabel~HN (RLHN)}: relabeling the false hard negatives as ground truth. 
We experiment on the seven pruned training datasets from the BGE training collection~\cite{li:2024c}. 

Our results consistently show that the RLHN setting achieves the highest nDCG@10 scores on \beir~\cite{thakur:2021} and \ab~\cite{chen-etal-2025-air}, amongst their counterparts with both retrievers:\ E5 (base) and Qwen2.5-7B and a reranker with Qwen2.5-3B. 
Compared to the aforementioned works,
RLHN outperforms hard negative sampling in \citet{moreira:2024} and is comparable to cross-encoder distillation~\cite{qu-etal-2021-rocketqa} yet with a simpler training pipeline.

To better understand the behavior of LLM judgment in identifying false negatives,
we compare LLM judgment with human assessors on 670 randomly sampled query--hard negative pairs.
We observe the Cohen's Kappa ($\kappa$) score of \gpt is 10 points higher than \gptmini,
which echoes their effectiveness in improving training data quality. 
Lastly, we provide a qualitative analysis examining different categories of false negatives identified in training datasets.
 
Our contributions are as follows:\
(1) We are the first to report that carelessly adopting enormous training data may negatively affect the retriever and reranker model training. We show that the retrieval effectiveness can be improved by 4\% with 57\% {\it less} data, 
(2) We propose a LLM cascading framework that identifies and relabels the false hard negatives at an instance level. Our approach results in higher in-domain and out-of-domain retrieval effectiveness with a simpler training pipeline. 

\section{Related Work}

\paragraph{Sparsely-annotated datasets.} Popular IR training datasets, such as \msmarco~\cite{nguyen:2016}, were shallow pooled and sparsely judged by human assessors~\cite{mackenzie:2021, arabzadeh:2022}. 
The assessor observed a few passages from a baseline retrieval system, picked those relevant to the query, and labeled them as ground-truth. 
On the other hand, non-relevant judged passages (i.e., passages seen but preferred lower than the ground truth) were not provided. 
Therefore, an assumption is made in fine-tuning where remaining passages (in a passage corpus) are negatives, and a few mined passages similar to the query are labeled as hard negatives. In this work, we avoid relabeling false positives, as these labels are trustworthy, provided by a human assessor, who can have a different preference than the LLM itself. 

\paragraph{LLM-based data curation.} Hiring human assessors for judgments is expensive and time-consuming, and produces limited training pairs, e.g., 1K pairs in LIMA~\cite{zhou:2023}. Alternatively, LLMs as judges have been recently explored for dataset curation in tasks, such as reranking~\cite{ma:2023, zhuang:2024, qin-etal-2024-large}, instruction fine-tuning~\cite{chen2024alpagasus, chen:2024b}, or even code-generation~\cite{jain2024llmassisted}.

\paragraph{Pseudo-labeling.} Instead of using supervised judgments, pseudo-labeling tackles the problem of sparse annotations by employing other techniques to estimate query-document relevance. Examples include distillation from cross-encoders~\cite{qu-etal-2021-rocketqa}, or ranking documents through prompting LLMs~\cite{sun:2023}, or through composite measures of embedding similarity with ground-truth documents~\cite{zerveas-etal-2023-enhancing}.

\paragraph{False negatives.} 
\citet{qu-etal-2021-rocketqa} first noted the issue of false negatives in retrieval, where certain hard negative passages should have been classified as positives. 
However, instead of \emph{curating} the training datasets, RocketQA~\cite{qu-etal-2021-rocketqa} fine-tuned models by distilling knowledge from the cross-encoder score for the query--document pair. 
Similarly, \citet{moreira:2024} examined various filtering methods for negative sampling by avoiding very hard negatives. 
In Gecko~\citep{lee:2024, lee:2025}, an LLM such as Gemini was used to relabel positive passages and identify better hard negatives. However, unlike our work, they focused on relabeling synthetic queries rather than existing collections like MS MARCO or NQ.

\section{The RLHN Methodology}
\label{sec:rlhn_methodology}
In this section, we discuss the LLM judge cascading framework, training dataset modifications, and dataset postprocessing and statistics.

\subsection{LLM Judge Cascading Framework}
\label{sec:rlhn_methodology:framework}
We adopt a simple and cost-effective approach of using cascaded LLM judges (shown in \autoref{fig:diagram}) inspired by \citet{chen:2023} to identify false hard negatives datasets at a large scale. The framework involves two stages:\

\begin{enumerate}[leftmargin=0.5cm, topsep=1pt]
    \item \textbf{Cost-effective judge (\gptmini):} In the first stage, we prompt \gptmini~\cite{openai_gpt4o}, a cost-effective LLM in the first stage to improve recall by identifying potential pairs with false negatives across all training pairs. 
    \item \textbf{Accurate judge (\gpt):} In the second stage, we prompt \gpt~\cite{openai_gpt4o}, a more reliable and expensive judge\footnote{\gptmini and \gpt pricing (as of May 15th, 2025) is 0.6\$ and 5.0\$ for 1M input tokens and 2.4\$ and 20.0\$ for 1M output tokens, respectively.} to re-evaluate the potential pairs with false negatives identified by \gptmini and re-evaluate them using \gpt to improve precision.
\end{enumerate}

\subsection{Training Dataset Modification}
\label{sec:rlhn_methodology:modification}

Upon successful completion of identifying the false negatives, we compare three operations on the identified false negatives as follows:

\begin{itemize}[leftmargin=0.3cm]

\item \textbf{Remove}: Discard the complete training instance due to the low quality, even if it contains at least one false negative.\footnote{We lose the instance completely in the ``remove'' technique.}

\item \textbf{Remove HN}: Discard only the detected false negatives from the hard negative subset, keeping the instance with the remaining hard negatives.

\item \textbf{Relabel HN (RLHN)}: Relabel only the detected false negatives from the hard negative subset, by adding them to the ground truth subset, keeping the instance with the remaining hard negatives. 

\end{itemize}

\begin{table}[t]
\centering
\resizebox{0.48\textwidth}{!}{
\begin{tabular}{lrrr|rr}
\toprule
\multirow{2}{*}{\textbf{Dataset}} & \textbf{\#Train} & \multicolumn{1}{c}{\textbf{Avg.}} & \multicolumn{1}{c|}{\textbf{Avg.}} & 
\multicolumn{2}{c}{\textbf{RLHN}} \\ 
& \textbf{Pairs} & \textbf{GT/Q} & \multicolumn{1}{c|}{\textbf{HN/Q}} & \textbf{Stage 1} & \textbf{Stage 2} \\
\midrule
\msmarco & 485,823 & 1.1 & 25.0 & 391,965 & 326,301 \\
\hotpotqa & 84,516 & 2.0 & 20.0 & 11,268 & 4,756 \\
\nq & 58,568 & 1.0 & 98.5 & 32,184 & 19,199 \\
\fever & 29,096 & 1.3 & 20.0 & 7,764 & 3,577 \\
\scidocsrr & 12,655 & 1.6 & 19.7 & 2,068 & 351 \\
\fiqa-2018 & 5,500 & 2.6 & 15.0 & 3,632 & 1,833 \\
\arguana & 4,065 & 1.0 & 13.6 & 0  &  0  \\
\bottomrule
\end{tabular}}
\caption{BGE training dataset statistics~\cite{chen:2024}. Avg.~GT/Q denotes the average ground truth passages per query, and Avg.~HN/Q denotes the average hard negative passages per query. RLHN Stages 1 \& 2 show training pairs with at least one false hard negative.}
\label{tab:train_stats}
\end{table}

\subsection{Dataset Postprocessing \& Statistics} 
In \autoref{tab:train_stats}, we show the training dataset statistics observed in the BGE training collection. \msmarco contains the highest amount of training pairs, followed by \hotpotqa. All datasets contain training pairs with 1--3 ground-truth passages and 13--25 hard negatives (except \nq with 98--100 hard negatives).  

\begin{figure}[t]
\centering
\begin{center}
\vspace{-0.4em}
    \includegraphics[trim=0 9 0 6,clip,width=0.48\textwidth]{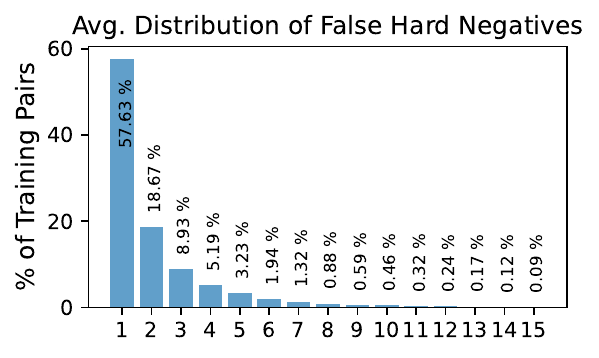}
    \caption{The distribution of training pairs (with at least one false negative) across false hard negatives detected. 58\% of the training pairs detected contain a single false negative, 19\% with two false negatives, and so on.}
    \label{fig:false-negatives}
\end{center}
\vspace*{-\baselineskip}
\end{figure}

\paragraph{False negatives.} From \autoref{tab:train_stats}, we see a majority of detected false negatives occur in \msmarco (91.6\% of all detected pairs). A maximum of up to 56\% of all training pairs in \msmarco contain false negatives, to a minimum of about 3\% in \scidocsrr.\footnote{We avoid relabeling \arguana due to its inherent complex task, which doesn't measure directly for argument similarity, but rather counter arguments given an argument. Therefore, we keep the original dataset in fine-tuning without relabeling.} From \autoref{fig:false-negatives}, we observe that in 58\% of all detected false negative pairs, only a single false positive was detected, and 19\% with two false negatives, and less than 1\% with eight or more false negatives.
If we detect any training pair with detected false negatives over a certain threshold $k$ ($k=7$ in our experiments), we excluded the pair completely in RLHN, as the query is likely to be ambiguous, that might not be a useful training instance (e.g., \emph{what color is amber urine?}).

\paragraph{Cost estimates.} We report the maximum costs incurred in RLHN (accurate input tokens + estimated 2048 output tokens on average) by both judges at each cascading stage: GPT-4o-mini and GPT-4o in \autoref{tab:cost_estimates}. 
Overall, running RLHN with \gptmini in Stage 1 costs around $\approx$~300 USD and with \gpt in Stage 2 costs around $\approx$~3000 USD.

\section{Experimental Setting}

\paragraph{BGE training data.} We utilize the original BGE training dataset\footnote{\href{https://huggingface.co/datasets/cfli/bge-full-data}{\texttt{huggingface.co/datasets/cfli/bge-full-data}}}~\cite{li:2024c}, a comprehensive collection with training datasets for retrieval (e.g., NQ, MS MARCO), clustering (e.g., TwentyNewsgroups), and classification (e.g., AmazonReviews) tasks. 
Many of these training datasets are used in fine-tuning of popular retriever models such as E5-Mistral~\cite{wang:2024b}, GRIT-LM~\cite{muennighoff2025generative}, Linq~\cite{choi:2024}, LLM2Vec~\cite{behnamghader2024llmvec}, CDE~\cite{morris2025contextual}, or NV-Embed~\cite{lee2025nvembed}. 
Our work focuses on the \emph{retrieval task}, therefore, we remove all training datasets from clustering and classification tasks, resulting in 15 datasets focused on the retrieval task, comprising a total of 1.6M training pairs, originally released with the MIT license.

\paragraph{LLM judges.} In our work, we use \gptmini (version \texttt{2024-07-18}) and \gpt (version \texttt{2024-11-20}) as the judge using the Azure OpenAI service in the \emph{batch} setting. 
We follow a temperature setting of $0.1$ and use a chain-of-thought prompt setting \cite{wei:2023}. The prompt first evaluates the relevance between every hard negative passage and the question, and compares them with the ground truth to identify potential false negatives. We prompt up to 25 hard negative passages per query in a single API call as shown in \autoref{fig:prompt}.

\paragraph{Evaluation benchmarks.} We evaluate the retrieval and reranker accuracy of the models fine-tuned on datasets with false negatives either removed or relabeled with RLHN on the \beir benchmark~\cite{thakur:2021} and \ab~\cite{chen-etal-2025-air}. 
Both benchmarks evaluate retrieval accuracy in nDCG@10. \beir contains human-constructed datasets, and \ab contains datasets automatically generated by LLMs without human intervention.
In \beir, we drop Quora and CQADupstack and evaluate on the remaining 16 datasets. In \ab (version \texttt{24.05}), we evaluate five specific domains in English-only:\ Arxiv, Finance, Healthcare, Law, and News.

\begin{table}[t]
\centering
\resizebox{0.48\textwidth}{!}{
\begin{tabular}{lrr|rr}
\toprule
& \multicolumn{2}{c}{\emph{Cascading Stage 1}} & \multicolumn{2}{c}{\emph{Cascading Stage 2}} \\ \cmidrule(lr){2-3} \cmidrule(lr){4-5}
\
\textbf{Dataset} & \textbf{\#~Pairs} & \textbf{GPT-4o-mini} & \textbf{\#~Pairs} & \textbf{GPT-4o} \\
\midrule
\msmarco & 485,823 & 180.40 USD & 391,965 & 2431.98 USD \\
\hotpotqa & 84,516 & 43.35 USD & 11,268 & 97.26 USD \\
\nq & 58,568 & 37.41 USD & 32,184 & 345.08 USD \\
\fever & 29,096 & 22.67 USD & 7,764 & 103.99 USD \\
\scidocsrr & 12,655 & 9.07 USD & 2,068 & 24.81 USD \\
\fiqa-2018 & 5,500 & 3.60 USD & 3,632 & 40.17 USD \\
\midrule
\textbf{Total Costs} &  & \textbf{\textasciitilde300 USD} &  & \textbf{\textasciitilde3000 USD} \\
\bottomrule
\end{tabular}}
\caption{Cost estimates for relabeling false negatives in RLHN using GPT-4o-mini and GPT-4o.}
\label{tab:cost_estimates}
\end{table}

\paragraph{Backbone models.} 
We use the E5 (base) unsupervised\footnote{\href{https://huggingface.co/intfloat/e5-base-unsupervised}{\texttt{intfloat/e5-base-unsupervised}} on HuggingFace.} ~\cite{wang:2022, wang:2024b}, a BERT-based encoder, due to its high accuracy on \beir (preliminary results in \autoref{ap:backbone}), the inclusion of a pre-training stage, and lower training complexity. E5 (base) contains 110M parameters, 12 layers, and a 768 embedding dimension with mean pooling. 
Also, we use a LLM-based decoder model with Qwen2.5-7B model\footnote{\href{https://huggingface.co/Qwen/Qwen2.5-7B} {\texttt{Qwen/Qwen2.5-7B}} on HuggingFace.} \cite{qwen:2024} with 7.61B parameters, 28 layers, and a 3584 embedding dimension with the \texttt{[EOS]} token pooling as the retrieval models.
In addition, we use Qwen2.5-3B model~\cite{qwen:2024}\footnote{\href{https://huggingface.co/Qwen/Qwen2.5-3B}{\texttt{Qwen/Qwen2.5-3B}} on HuggingFace.} for the reranker.

\paragraph{Fine-tuning details.} 
All models were fine-tuned using 7 hard negatives, 1 positive, and random in-batch negatives (128 total) per batch, optimized with the InfoNCE loss function~\cite{oord:2018} using the Tevatron repository\footnote{ \href{https://github.com/texttron/tevatron}{\texttt{https://github.com/texttron/tevatron}}}~\cite{tevatron-v1, ma2025tevatron20unifieddocument} for up to 4–5 epochs, with a learning rate of 2e-5, and a maximum sequence length of 350 tokens (512 tokens during inference). We append a ``\texttt{query: }'' and ``\texttt{passage: }'' prefix. E5 (base) models are fine-tuned using 4$\times$L40S GPUs, and Qwen2.5-7B and Qwen2.5-3B using a maximum of 2$\times$H200 GPUs.

\begin{figure}[t]
\centering
\begin{center}
    \includegraphics[trim=0 10 0 0,clip,width=0.48\textwidth]{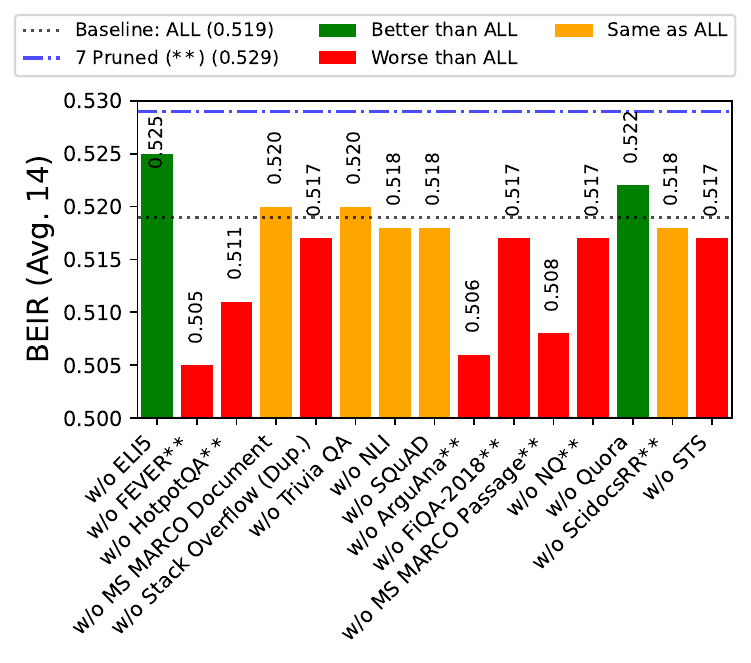}
    \caption{Dataset pruning by leaving one dataset out during fine-tuning E5 (base) on the BGE-training collection; [ALL] denotes fine-tuning on all datasets with 1.6M training pairs; [7 Pruned] denotes fine-tuning on 680K training pairs with seven remaining datasets (or 57.5\% pairs) after dataset pruning.
    [Better than ALL] denotes the results \textit{improved} after \textit{removing} the dataset, meaning it has negative impact on the training process.
    [Worse than ALL] denotes the opposite, where the dataset has a positive impact on the training.
    }
    \label{fig:beir-leave-one-out-scores}
\end{center}
\vspace*{-\baselineskip}
\end{figure}

\paragraph{Baselines.} To evaluate the impact of relabeling hard negatives using RLHN, we include two baselines: (1) \emph{hard-negative mining}: Top-95\% TopK-PecPos sampling~\cite{moreira:2024} on the default training dataset, using similarity scores computed for all hard negatives with the \texttt{bge-reranker-v2-gemma} reranker, and (2) \emph{cross-encoder distillation}: we compute the normalized similarity scores for all query and hard negatives and positive pair on the default training dataset with the \texttt{bge-reranker-v2-gemma} reranker. We fine-tune the E5-base using knowledge distillation from the cross-encoder scores, with 1 positive, 15 hard and zero in-batch negatives using Tevatron.

\begin{table*}[!ht]
\centering
\resizebox{\textwidth}{!}{
\begin{tabular}{l|ccc|ccc|ccc|c|cc}
\toprule
\multirow{2}{*}{\textbf{BEIR Dataset}} & \multicolumn{1}{c}{\emph{No Filtering}} & \multicolumn{2}{c}{\emph{Baselines}}  & \multicolumn{3}{c}{\emph{Cascading Stage 1: \gptmini}} & \multicolumn{3}{c}{\emph{Cascading Stage 2: \gptmini + \gpt}} & \multicolumn{1}{c}{\emph{No Filtering}} & \multicolumn{2}{c}{\emph{Cascading Stage 2}} \\ \cmidrule(lr){2-2} \cmidrule(lr){3-4} \cmidrule(lr){5-7} \cmidrule(lr){8-10} \cmidrule(lr){11-11} \cmidrule(lr){12-13}

& \textbf{Default} & \textbf{TopK-PercPos} & \textbf{CE Distill} & \textbf{Remove} & \textbf{Remove HN} & \textbf{RLHN} & \textbf{Remove} & \textbf{Remove HN} & \textbf{RLHN} & \textbf{Default} & \textbf{Remove HN} & \textbf{RLHN} \\ 
\midrule
\textbf{Backbone}      & E5 (base) & E5 (base)  & E5 (base) & E5 (base) & E5 (base) & E5 (base) & E5 (base) & E5 (base) & E5 (base) & Qwen2.5-7B & Qwen2.5-7B & Qwen2.5-7B \\ \midrule

\textbf{TREC-COVID}$^\dagger$     & 0.783 & 0.789 & 0.793 & 0.786 & 0.793 & 0.798 & 0.794 & 0.785 & 0.809 & 0.797 & 0.771 & 0.815 \\
\textbf{NFCorpus}$^\dagger$       & 0.378 & 0.377 & 0.363 & 0.378 & 0.380 & 0.381 & 0.380 & 0.382 & 0.390 & 0.389 & 0.389 & 0.391 \\
\textbf{NQ}                       & 0.595 & 0.601 & 0.624 & 0.593 & 0.592 & 0.602 & 0.573 & 0.598 & 0.591 & 0.597 & 0.602 & 0.623 \\
\textbf{HotpotQA}                 & 0.737 & 0.734 & 0.741 & 0.737 & 0.736 & 0.739 & 0.741 & 0.736 & 0.735 & 0.704 & 0.702 & 0.729 \\
\textbf{FiQA-2018}                & 0.439 & 0.434 & 0.417 & 0.443 & 0.440 & 0.444 & 0.441 & 0.445 & 0.448 & 0.453 & 0.461 & 0.465 \\
\textbf{ArguAna}                  & 0.701 & 0.697 & 0.725 & 0.702 & 0.706 & 0.700 & 0.700 & 0.700 & 0.692 & 0.554 & 0.550 & 0.560 \\
\textbf{Touch\'e-2020}$^\dagger$  & 0.256 & 0.286 & 0.305 & 0.255 & 0.271 & 0.268 & 0.218 & 0.265 & 0.266 & 0.221 & 0.211 & 0.230 \\
\textbf{DBPedia}                  & 0.438 & 0.444 & 0.446 & 0.439 & 0.437 & 0.442 & 0.433 & 0.441 & 0.447 & 0.443 & 0.456 & 0.472 \\
\textbf{SCIDOCS}                  & 0.242 & 0.243 & 0.216 & 0.243 & 0.243 & 0.244 & 0.245 & 0.243 & 0.242 & 0.245 & 0.243 & 0.252 \\
\textbf{FEVER}                    & 0.878 & 0.878 & 0.889 & 0.875 & 0.876 & 0.877 & 0.881 & 0.876 & 0.871 & 0.863 & 0.857 & 0.872 \\
\textbf{Climate-FEVER}            & 0.391 & 0.386 & 0.377 & 0.388 & 0.385 & 0.391 & 0.382 & 0.384 & 0.367 & 0.370 & 0.373 & 0.360 \\
\textbf{SciFact}                  & 0.735 & 0.735 & 0.727 & 0.741 & 0.731 & 0.733 & 0.744 & 0.735 & 0.740 & 0.755 & 0.755 & 0.767 \\
\textbf{TREC-NEWS}$^\dagger$      & 0.465 & 0.466 & 0.458 & 0.470 & 0.466 & 0.473 & 0.464 & 0.473 & 0.484 & 0.494 & 0.480 & 0.487 \\
\textbf{Robust04}$^\dagger$       & 0.442 & 0.451 & 0.452 & 0.448 & 0.452 & 0.471 & 0.447 & 0.458 & 0.497 & 0.501 & 0.501 & 0.540 \\
\textbf{Signal-1M (RT)}$^\dagger$ & 0.275 & 0.272 & 0.271 & 0.279 & 0.275 & 0.275 & 0.274 & 0.270 & 0.274 & 0.275 & 0.268 & 0.280 \\
\textbf{BioASQ}$^\dagger$         & 0.378 & 0.375 & 0.413 & 0.382 & 0.385 & 0.392 & 0.384 & 0.384 & 0.394 & 0.408 & 0.412 & 0.438 \\ \midrule
\textbf{Avg. 16 (All)}            & 0.508 & 0.511 & 0.514 & 0.510 & 0.511 & 0.514 & 0.506 & 0.511 & \textbf{0.515} & 0.504 &  0.502 & \textbf{0.518} \\
\textbf{Avg. 7 (OOD)}             & 0.425 & 0.431 & 0.436 & 0.428 & 0.432 & 0.437 & 0.423 & 0.431 & \textbf{0.445} & 0.441 & 0.433 & \textbf{0.454} \\
\bottomrule
\end{tabular}}
\vspace{-0.5em}
\caption{Retrieval results measuring nDCG@10 on 16 datasets in the BEIR benchmark by fine-tuning retrieval models on variants of the BGE training dataset after relabeling false negatives. The seven unseen (or out-of-domain) datasets during fine-tuning are highlighted with $\dagger$ and their average scores are provided in Avg. 7.}
\label{tab:transposed-beir-results}
\end{table*}

\section{Experimental Results}\label{sec:results}

\subsection{Preliminary Results: Dataset Pruning}
\paragraph{False datapoint can hurt the training of retriever models.} 
We assess the individual dataset contribution by evaluating several model variants by leaving one dataset out and fine-tuning the rest. As we fine-tune many models, i.e., one for each removed dataset, we limit these experiments to E5 (base).
Summarized results are shown in \autoref{fig:beir-leave-one-out-scores} (detailed results can be found in \autoref{tab:leave-one-out}), demonstrating that training datasets (highlighted in red) can hurt the model retrieval accuracy, such as ELI5, removing which improves the nDCG@10 on \beir (0.519 $\rightarrow$ 0.525). Also, it shows that certain datasets (highlighted in green) are crucial for model accuracy.

\noindent
Based on findings in \autoref{fig:beir-leave-one-out-scores} and selecting necessary datasets for individual task-based performances in \beir, we prune the original 16 retrieval datasets in the BGE collection and select seven datasets (highlighted as $**$), reducing the training dataset size from 1.6M to 680K training pairs in our experiments. The average nDCG@10 score of E5 (base) improves from 0.519 $\rightarrow$ 0.529 on 14 datasets on average in \beir, by fine-tuning on almost 2.35$\times$ smaller dataset (1.6M $\rightarrow$ 680K). 

\subsection{Main Results: Relabeling False Negatives}
 This section shows the results of the fine-tuned models on the variants of the training dataset described in Section~\ref{sec:rlhn_methodology:framework} and \ref{sec:rlhn_methodology:modification},
keeping the rest of the model training parameters unchanged.

\begin{table}[t]
\centering   
\resizebox{0.48\textwidth}{!}{%
\begin{tabular}{lrcrrrrr}
\toprule
\textbf{Backbone} & \textbf{Technique} & \textbf{Arxiv} & \textbf{Finance} & \textbf{Health.} & \textbf{Law} & \textbf{News} & \textbf{Avg. 5} \\ 
\midrule
E5 (base) & \textbf{Default} & 0.345 & 0.401 & 0.521 & 0.117 & 0.455 & 0.368 \\ 
E5 (base) & \textbf{TopK-PercPos} & 0.348 & 0.418 & 0.529 & 0.119 & 0.464 & 0.376 \\
E5 (base) & \textbf{CE Distill} & 0.372 & 0.430 & 0.536 & 0.168 & 0.498 & \textbf{0.401} \\

\rowcolor{paleaqua} \multicolumn{8}{c}{\emph{Cascading Stage 1: \gptmini}} \\ 
E5 (base) & \textbf{Remove}  & 0.346 & 0.407 & 0.526 & 0.118 & 0.452 & 0.370 \\
E5 (base) & \textbf{Remove HN} & 0.344 & 0.406 & 0.522 & 0.118 & 0.459 & 0.370 \\
E5 (base) & \textbf{RLHN} & 0.362 & 0.421 & 0.522 & 0.123 & 0.465 & \textbf{0.379} \\
\rowcolor{paleaqua} \multicolumn{8}{c}{\emph{Cascading Stage 2: \gptmini + \gpt}} \\ 

E5 (base) & \textbf{Remove} & 0.341 & 0.403 & 0.514 & 0.125 & 0.438 & 0.364  \\ 
E5 (base) & \textbf{Remove HN} & 0.346 & 0.411 & 0.525 & 0.124 & 0.464 & 0.374 \\
E5 (base) & \textbf{RLHN} & 0.356 & 0.440 & 0.521 & 0.138 & 0.476 & \textbf{0.386} \\ \midrule \midrule
Qwen2.5-7B & \textbf{Default} & 0.325 & 0.391 & 0.479 & 0.115 & 0.430 & 0.348 \\
\rowcolor{paleaqua} \multicolumn{8}{c}{\emph{Cascading Stage 2: \gptmini + \gpt}} \\ 
Qwen2.5-7B & \textbf{Remove HN} & 0.335 & 0.384 & 0.487	& 0.111	& 0.423 & 0.348 \\
Qwen2.5-7B & \textbf{RLHN} & 0.330 & 0.418 & 0.494 & 0.133 & 0.450 & \textbf{0.365} \\
\bottomrule
\end{tabular}}
\vspace{-0.5em}
\caption{
Retrieval results measuring nDCG@10 on five specialized domains in \ab dev (version \texttt{24.05}) by fine-tuning E5 (base) and Qwen2.5-7B on variants of the BGE training dataset with RLHN.}
\vspace{-0.5em}
\label{tab:air-bench}
\end{table}

\begin{table}[t]
\centering

\resizebox{\linewidth}{!}{
    \begin{tabular}{lccc}
    \toprule
    \multirow{2}{*}{\textbf{BEIR Dataset}}
     & \multicolumn{1}{c}{\textit{No Filtering}} 
     & \multicolumn{1}{c}{\textit{Cascading Stage 1}} 
     & \multicolumn{1}{c}{\textit{Cascading Stage 2}} 
     \\
     \cmidrule(lr){2-2} \cmidrule(lr){3-3} \cmidrule(lr){4-4}
     & \multicolumn{1}{c}{\textbf{Default}} & \multicolumn{1}{c}{\textbf{RLHN}} & \multicolumn{1}{c}{\textbf{RLHN}} \\
     \midrule
     
    \textbf{TREC-COVID}$^\dagger$ & 0.836 & 0.861 & 0.862 \\
    \textbf{NFCorpus}$^\dagger$ & 0.401 & 0.414 & 0.415 \\
    \textbf{NQ} & 0.730 & 0.739 & 0.736 \\
    \textbf{HotpotQA} & 0.863 & 0.861 & 0.861 \\
    \textbf{FiQA-2018} & 0.517 & 0.521 & 0.519 \\
    \textbf{ArguAna} & 0.740 & 0.730 & 0.763 \\
    \textbf{Touch\'e-2020}$^\dagger$ & 0.275 & 0.308 & 0.313 \\
    \textbf{DBPedia} & 0.532 & 0.536 & 0.538 \\
    \textbf{SCIDOCS} & 0.278 & 0.273 & 0.270 \\
    \textbf{FEVER} & 0.941 & 0.939 & 0.936 \\
    \textbf{Climate-FEVER} & 0.457 & 0.468 & 0.430 \\
    \textbf{SciFact} & 0.786 & 0.793 & 0.794 \\
    \textbf{TREC-NEWS}$^\dagger$ & 0.507 & 0.513 & 0.527 \\
    \textbf{Robust04}$^\dagger$ & 0.531 & 0.548 & 0.589 \\
    \textbf{Signal-1M}$^\dagger$ & 0.292 & 0.276 & 0.274 \\
    \textbf{BioASQ}$^\dagger$ & 0.510 & 0.505 & 0.500 \\
    \midrule
    \textbf{Avg. 16 (All)} & {0.575} & {0.580} & \textbf{0.583} \\
    \textbf{Avg. 7 (OOD)} & {0.479} & {0.489} & \textbf{0.497} \\
    \bottomrule
    \end{tabular}
}
\vspace{-0.5em}
\caption{
Reranker results measuring nDCG@10 on 16 datasets in BEIR by fine-tuning reranker models (based on Qwen2.5-3B) on variants of the BGE training datasets after relabeling false negatives.
Stage 1 and 2 refers to \gptmini and \gptmini + \gpt.
}
\label{tab:beir-reranker}
\vspace{-0.5em}
\end{table}
\paragraph{\beir benchmark.} Results in \autoref{tab:transposed-beir-results} show that for both E5 (base) and Qwen2.5-7B, the RLHN technique achieves the best overall average nDCG@10 of 0.515 and 0.518 on 16 datasets on \beir, outperforming models trained with the default setting and other remove techniques. 
The relabeled data in RLHN improves model generalization, with improvements strongly visible in seven out-of-domain (OOD) datasets in \beir. 
Stage 1 (RLHN) outperforms the Default setting by 2.0 points and Stage 2 (RLHN) by 3.2 points in nDCG@10. 
Overall, relabeling false negatives improves the data quality, which is reflected in model generalization across out-of-domain settings in \beir.

\paragraph{\ab.} In addition to \beir, \ab provides a zero-shot setting to evaluate on challenging domains, such as Law. 
\autoref{tab:air-bench} shows the average nDCG@10 on five specialized domains. The improvements in model generalization are consistent to what we observed in \beir.
Stage 1 (RLHN) improves the Default setting by 1.1 points in nDCG@10, and Stage 2 (RLHN) further improves by 2.1 points. Overall, without changing the model or training parameters, mitigating false negatives in training datasets with RLHN enables the model generalize better to specialized domains in \ab.

\paragraph{Comparison with baselines.} Results in \autoref{tab:transposed-beir-results} and \autoref{tab:air-bench} show that carefully avoiding sampling very hard negatives using Top-95\%-PercPos outperforms the Default model, but still  underperforms compared to the RLHN strategy with E5 (base). 
Next, the \texttt{bge-reranker-v2-gemma} cross-encoder, used as the distillation teacher is a strong baseline. It slightly underperforms RLHN on \beir but outperforms RLHN on \ab. However, we want to reiterate that our core motivation is to \emph{identify} and \emph{relabel} false negatives in training datasets to enhance data quality. Distillation-based fine-tuning requires on a strong, domain-focused cross-encoder reranker. Similarly, RLHN is particularly valuable for fine-tuning cross-encoders when teacher supervision is not viable.

\paragraph{Reranker results.} Training data with improved quality also benefits  cross-encoder rerankers.
\autoref{tab:beir-reranker} shows the result comparison on the \beir benchmark, where we rerank the top-100 results from the fine-tuned E5 (base) in the Default setting.
Training rerankers with data fixed on RLHN Stages 1 and 2 progressively increases nDCG@10 on \beir datasets by 0.5 points and 0.8 points, respectively.
This improvement is most prominent on the seven OOD datasets, consistent with the above observation on retrievers:\ the data correction on the two stages improves the averaged OOD results by 1.0 and 1.8 points, respectively. 

We note that the scale of the improvement on cross-encoders is not as large as on retrievers,
which may indicate that cross-encoder rerankers are comparatively more robust to false negatives.
However, albeit small, cross-encoders still benefit from training data of higher quality, especially when generalizing to unseen domains. 

\begin{table}[t]
\centering

\resizebox{\linewidth}{!}{
    \begin{tabular}{lccccc}
    \toprule
    \multirow{2}{*}{\textbf{BEIR Dataset}}
     & \multicolumn{4}{c}{\textit{RLHN (Ablation of Hard Negatives)}} \\
     \cmidrule(lr){2-5}
      & \textbf{RLHN (1 HN)} & \textbf{RLHN (3 HN)} & \textbf{RLHN (7 HN)} & \textbf{RLHN (9 HN)} \\
     \midrule
     
    \textbf{TREC-COVID}$^\dagger$ & 0.809 & 0.810 & 0.809 & 0.812 \\
    \textbf{NFCorpus}$^\dagger$ & 0.389 & 0.388 & 0.390 & 0.392 \\
    \textbf{NQ}  & 0.563 & 0.583 & 0.591 & 0.595 \\
    \textbf{HotpotQA} & 0.717 & 0.729 & 0.735 & 0.739 \\
    \textbf{FiQA-2018} & 0.438 & 0.448 & 0.448 & 0.450 \\
    \textbf{ArguAna} & 0.660 & 0.679 & 0.692 & 0.693 \\
    \textbf{Touch\'e-2020}$^\dagger$ & 0.249 & 0.263 & 0.266 & 0.276 \\
    \textbf{DBPedia} & 0.439 & 0.442 & 0.447 & 0.447 \\
    \textbf{SCIDOCS} & 0.234 & 0.238 & 0.242 & 0.243 \\
    \textbf{FEVER}  & 0.851 & 0.864 & 0.871 & 0.875 \\
    \textbf{Climate-FEVER}  & 0.339 & 0.362 & 0.367 & 0.371 \\
    \textbf{SciFact}  & 0.736 & 0.737 & 0.740 & 0.744 \\
    \textbf{TREC-NEWS}$^\dagger$  & 0.481 & 0.473 & 0.484 & 0.484 \\
    \textbf{Robust04}$^\dagger$  & 0.506 & 0.502 & 0.497 & 0.499 \\
    \textbf{Signal-1M}$^\dagger$  & 0.273 & 0.272 & 0.274 & 0.272 \\
    \textbf{BioASQ}$^\dagger$  & 0.384 & 0.394 & 0.394 & 0.397 \\
    \midrule
    \textbf{Avg. 16 (All)} & {0.504} & {0.512} & {0.515} & \textbf{0.518} \\
    \textbf{Avg. 7 (OOD)} & {0.442} & {0.443} & {0.445} & \textbf{0.447} \\
    \bottomrule
    \end{tabular}
}
\vspace{-0.5em}
\caption{
Ablation of number of hard negatives during fine-tuning with InfoNCE loss function~\cite{oord:2018} in Tevatron with E5 (base).
}
\label{tab:rlhn_ablation}
\vspace{-0.5em}
\end{table}

\begin{figure*}[t]
\centering
\begin{center}
    \includegraphics[trim=0 5 0 0,clip,width=0.8\textwidth]{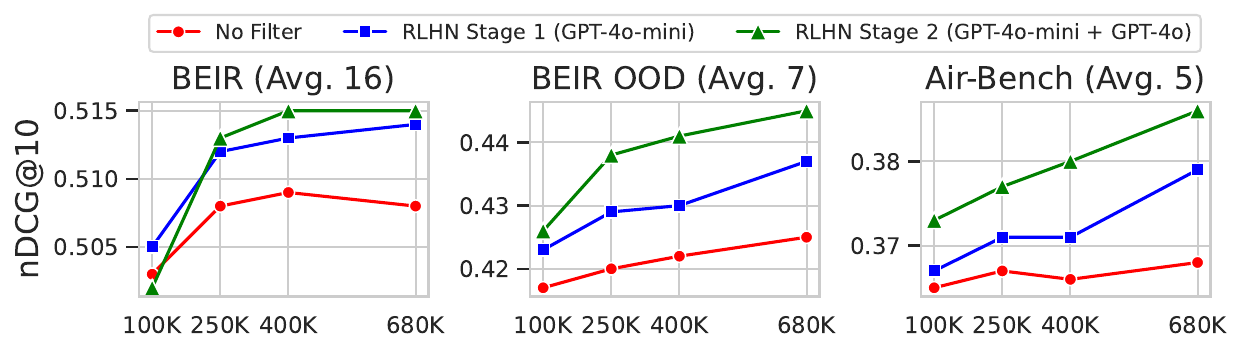}
    \caption{nDCG@10 scores on \beir (Avg. 16 and Avg. 7) and \ab (Avg. 5) by fine-tuning E5 (base) on a subset of the 100K, 250K, 400K, and 680K training pairs using the ``RLHN'' technique for both stages. All individual dataset scores for both \beir and \ab are provided in \autoref{fig:all-scores-beir} and \autoref{fig:all-scores-air-bench}.}
    \label{fig:training-pairs-ablation}
\end{center}
\vspace*{-1em}
\end{figure*}

\begin{table*}[!ht]
\centering
\resizebox{\textwidth}{!}{
\begin{tabular}{l|cc|cc|cc|cc}
\toprule
\textbf{Datasets $\rightarrow$} & \multicolumn{2}{c}{\emph{\fever} (3,521)} & \multicolumn{2}{c}{\fiqa-2018 (1,829)} & \multicolumn{2}{c}{\emph{\hotpotqa} (4,720)} &  \multicolumn{2}{c}{\emph{\scidocsrr} (350)} \\ \cmidrule(lr){2-3} \cmidrule(lr){4-5} \cmidrule(lr){6-7} \cmidrule(lr){8-9} 
\textbf{SoTA Reranker Judge $\downarrow$} & \textbf{mAP@10} & \textbf{P@L(GT)} & \textbf{mAP@10} & \textbf{P@L(GT)} & \textbf{mAP@10} & \textbf{P@L(GT)} & \textbf{mAP@10} & \textbf{P@L(GT)} \\ 
\midrule
 \texttt{BAAI/bge-reranker-v2-gemma}               & \textbf{0.839} & \textbf{0.777} & 0.632 & 0.492 & \textbf{0.742} & \textbf{0.638} & \textbf{0.926} & \textbf{0.875} \\
 \texttt{mxbai/rerank-large-v2}                    & 0.496 & 0.365 & \textbf{0.658} & \textbf{0.525} & 0.737 & 0.634 & 0.680 & 0.524 \\
 \texttt{mxbai/rerank-base-v2}                     & 0.570 & 0.455 & 0.598 & 0.464 & 0.671 & 0.565 & 0.612 & 0.462 \\
 \texttt{Cohere (rerank-v3.5)}                     & 0.811 & 0.740 & 0.572 & 0.437 & 0.694 & 0.588 & 0.838 & 0.743 \\
 \texttt{Alibaba-NLP/gte-reranker-modernbert-base} & 0.688 & 0.602 & 0.545 & 0.408 & 0.658 & 0.560 & 0.843 & 0.754 \\
 \texttt{cross-encoder/ms-marco-MiniLM-L12-v2}     & 0.745 & 0.656 & 0.517 & 0.387 & 0.587 & 0.479 & 0.832 & 0.755 \\
\bottomrule
\end{tabular}}
\vspace{-0.5em}
\caption{Reranker as the judge as a baseline to identify RLHN false negatives in each training dataset (written along with the count of training pairs). \textbf{mAP@10} calculates the average precision of false negatives (labeled as positives) in the top-10 reranked results. \textbf{P@L(GT)} calculates the precision of false negatives present in top-$k$ reranked results, where $k$ varies in each query, measuring the count of false negatives detected using RLHN.}
\label{tab:reranker-judge-results}
\end{table*}

\section{Analysis}

\paragraph{Ablation on hard-negatives and significance tests.} As an ablation, we experiment with the number of hard negatives during fine-tuning E5 (base) in Tevatron. From \autoref{tab:rlhn_ablation}, we observe that increasing the number of hard negatives improves the nDCG@10 score on \beir, with the best scores observed using 9 hard negatives. 

We conduct statistical significance tests using ranger plots~\cite{sertkan-etal-2023-ranger} for both E5 (base) and Qwen2.5-7B, comparing RLHN versus the Default setting. The ranger plots are provided in the Appendix (\autoref{fig:ranger-plot-e5-base} and \autoref{fig:ranger-plot-qwen-base}). In \autoref{fig:ranger-plot-e5-base}, the plot shows statistical improvement for 10/16 \beir datasets with E5 (base) fine-tuned using RLHN. Similarly,  \autoref{fig:ranger-plot-qwen-base} shows statistical improvement for 14/16 \beir datasets with Qwen2.5-7B fine-tuned using RLHN.

\paragraph{Robustness of RLHN across varying training data subsets.} As training datasets can be large, relabeling all training pairs using the LLM cascading pipeline can be computationally prohibitive. 
From \autoref{fig:training-pairs-ablation}, we demonstrate that RLHN remains robust and maintains similar accuracy gains, even when applied to smaller randomly sampled subsets of the training dataset. 
To evaluate this, we use four random subsets (100K, 250K, 400K, and 680K) of the training datasets, with each dataset's distribution shown in \autoref{tab:label-distribution}.

Overall, we have two main findings: (i) the E5 (base) model fine-tuned on RHLN Stages 1 and 2 training data, with false hard negatives relabeled as positives, \emph{consistently} outperforms the Default setting, and (ii) the steeper slope in nDCG@10 demonstrates \emph{continual improvement} across zero-shot domains, as the amount of training data increases, especially as observed in \ab.

\paragraph{Reranker distillation is competitive but limited in detecting false negatives.}
A reranker, or cross-encoder, is commonly used in knowledge distillation to fine-tune a retriever model as an alternative to the traditional contrastive or InfoNCE loss function~\cite{hofstatter:2020, qu-etal-2021-rocketqa, wang-etal-2022-gpl}. 
This approach bypasses the original relevance judgments, relying instead on knowledge encoded within the reranker itself.
Rather than using RLHN, we evaluate how well rerankers detect false negatives in training datasets. 
Specifically, we rerank the hard negatives for each training instance and compute two metrics: (i) mAP@10, which measures the average precision of false negatives in the top-10 results, and (ii) P@L(GT), which measures the precision of false negatives among the top-$k$ results, where $k$ equals the number of false negatives. 

\autoref{tab:reranker-judge-results} reports results of six reranker judges from various sources across four datasets. 
We observe that the \texttt{bge-reranker-v2-gemma} judge achieves the highest scores amongst its counterparts in identifying false negatives labeled by RLHN (except on \fiqa). 
However, on datasets such as \fiqa-2018 and \hotpotqa, rerankers detect only 52.5--63.8\% of false negatives, indicating that while existing rerankers are competitive, they still require improvement. 
We suspect this limitation arises because rerankers are fine-tuned on these existing training datasets that contain false negatives, which negatively affects their accuracy.

\begin{table}[t]
    \centering
    \resizebox{\linewidth}{!}{ 
    \begin{tabular}{ccc}
    \toprule
        \textbf{Metric} & ~~~~\textbf{\gptmini}~~~~ & ~~~~\textbf{\gpt}~~~~ \\ 
    \midrule
         Cohen's Kappa ($\kappa)$ & 0.320 & 0.390 \\
    \bottomrule
    \end{tabular}
    }
    \vspace{-0.5em}
    \caption{Cohen's $\kappa$ score of \gptmini and \gpt with human judgments on 670 query--negative pairs.}
    \label{tab:human_validation}
    \vspace{-0.5em}
\end{table}

\begin{table*}[!ht]
\centering
\resizebox{\textwidth}{!}{
\begin{tabular}{c|cc|cc}
\toprule
\multirow{1}{*}{\textbf{Query}} & \multicolumn{2}{c}{\textbf{Ground Truth or Positive Passages}} & \multicolumn{2}{c}{\textbf{False Negatives (Detected by RLHN)}} \\
 \cmidrule(lr){1-1} \cmidrule(lr){2-3} \cmidrule(lr){4-5} 

\multicolumn{1}{p{3cm}|}{(Q1) Which is a food magazine, Latin Mass Magazine or Saveur?} & \multicolumn{1}{p{7cm}}{\textbf{Latin Mass Magazine}: \hlgreen{A Journal of Catholic Culture, commonly referred to as Latin Mass Magazine, is an American Catholic magazine published quarterly}, with a traditionalist Catholic viewpoint. [ ... ]} &
\multicolumn{1}{p{7cm}|}{\textbf{Saveur}: \hlgreen{Saveur is a gourmet, food, wine, and travel magazine that specializes in essays about various world cuisines.} Its slogan—"Savor a World of Authentic Cuisine"—signals the publication's focus on enduring culinary traditions [ ... ]} & 
\multicolumn{1}{p{7cm}}{\textbf{Food \& Wine}: \hlred{Food \& Wine is a monthly magazine published by Time Inc. It was founded in 1978 by Ariane and Michael Batterberry. It features recipes, cooking tips, travel information, restaurant reviews, chefs, wine pairings and seasonal content [ ... ]} } &
\multicolumn{1}{p{7cm}}{\textbf{Cocina (magazine)}: \hlred{is a Colombian-based monthly magazine published by Publicaciones Semana S.A.. It features recipes, cooking tips, culinary tourism information, restaurant reviews, chefs, wine pairings and seasonal holiday content [ ... ] }} \\ \midrule

\multicolumn{1}{p{3cm}|}{(Q2) What year was the premier professional ice hockey league in the world established?} & \multicolumn{1}{p{7cm}}{\textbf{2016–17 Minnesota Wild season}: The 2016–17 Minnesota Wild season was the \hlred{17th season for the National Hockey League franchise that was established on June 25, 1997.}} &
\multicolumn{1}{p{7cm}|}{\textbf{National Hockey League}: \hlred{The National Hockey League (NHL; French: "Ligue nationale de hockey—LNH" ) is a professional ice hockey league currently comprising 31 teams} [ ... ]} & 
\multicolumn{2}{p{14.5cm}}{\textbf{History of the National Hockey League (1917–42)}: History of the National Hockey League (1917–42) \hlgreen{The National Hockey League (NHL) was founded in 1917 following the demise of its predecessor league, the National Hockey Association (NHA).} [ ... ] } \\ \midrule

\multicolumn{1}{p{3cm}|}{(Q3) name meaning  yin and yang} & \multicolumn{2}{p{14.5cm}|}{\textbf{Yin and yang}: In Chinese philosophy, \hlgreen{yin and yang (also, yin-yang or yin yang) describes how apparently opposite or contrary forces are actually complementary, interconnected, and interdependent in the natural world}, and how they give rise to each other as they interrelate to one another.} & \multicolumn{1}{p{7cm}}{\textbf{Yin and yang}: \hlgreen{Yin and Yang are ancient Chinese philosophical terms, with the Yin Yang Theory being a fundamental part of Feng Shui. It is a Chinese theory on the perspective of continuous change and balance.} [ ... ] } & \multicolumn{1}{p{7cm}}{\textbf{Yin Yang Symbols and Their Meanings}: In a nutshell, \hlgreen{Chinese yin yang symbols represent perfect balance. A great deal of Chinese philosophy stems from the concept of yin and yang - opposites interacting} [ ... ] } \\ \midrule

\multicolumn{1}{p{3cm}|}{(Q4) Charles, Prince of Wales is patron of numerous other organizations.} & \multicolumn{2}{p{14.5cm}|}{\textbf{Charles, Prince of Wales}: Charles, Prince of Wales (born 14 November 1948) is the eldest child and heir apparent of Queen Elizabeth II [ ... ] \hlgreen{Charles's interests encompass a range of humanitarian and social issues: he founded The Prince's Trust in 1976, sponsors The Prince's Charities, and is patron of numerous other charitable and arts organisations.} [ ... ]} & \multicolumn{1}{p{7cm}}{\textbf{Julia Cleverdon Dame}: Julia Charity Cleverdon [ ... ] served for 16 years as \hlorange{Chief Executive of Business in the Community, one of the Prince's Charities of Charles, Prince of Wales.}} & \multicolumn{1}{p{7cm}}{\textbf{The Prince's Trust}: \hlgreen{The Prince's Trust is a charity in the United Kingdom founded in 1976 by Charles, Prince of Wales, and Frederick John Pervin to help young people.} [ ... ] } \\ 
\bottomrule
\end{tabular}}
\caption{Qualitative analysis showcasing the different varieties of false negatives detected by RLHN. The first two questions are taken from \hotpotqa, the third from \msmarco, and the last from \fever. The text supporting the query is highlighted in \hlgreen{green}, partially supporting in \hlorange{orange}, and not supporting with \hlred{red}.}
\label{tab:case-study}
\end{table*}

\section{Human Validation}

We conducted a validation study with three human assessors conducting using Label Studio\footnote{ \href{https://github.com/HumanSignal/label-studio/}{\texttt{github.com/HumanSignal/label-studio}}} for data annotation.
The assessors were briefed on the relevance task,
and then independently evaluated a total of 670 query--hard negative pairs.
The hard negatives were randomly sampled from the RLHN set, each containing at least one false negative. 
During the assessment,
all annotators worked independently and were not exposed to the LLM predictions.
An example of the annotation interface is shown in \autoref{fig:label-studio}.

\autoref{tab:human_validation} reports Cohen’s Kappa ($\kappa$) measuring agreement between each LLM’s predictions and the human labels. The $\kappa$ scores are consistent with prior work reporting similar levels of human--LLM agreement~\cite{arabzadeh2025human}. 
\gpt shows substantially higher agreement with human annotators compated to \gptmini.
This finding aligns with our empirical results, where relabeling with \gpt shows consistent gains over \gptmini in training  retrieval and reranker models.

\section{Qualitative Analysis of False Negatives}
We qualitatively analyze the labeling accuracy of our LLM cascading framework by manually spot-checking a few training instances. As shown in \autoref{tab:case-study}, we observe a variety of false negatives, which fall into the following scenarios:\

\paragraph{1. Detected false negatives are incorrect or not relevant.} GPT-4o can sometimes detect a false negative that is not relevant to the query. E.g., (Q1) query asks which is a food magazine between \emph{Latin Mass} or \emph{Saveur}, however, the detected false negatives identify different food magazines such as \emph{Food \& Wine} or \emph{Cochina}, which are both incorrect.

\paragraph{2. The ground truth may be incorrectly labeled.} In a few queries, we observe that the ground truth passage can contain conflicting information with the false negative, resulting in incorrect labeling. E.g., the correct answer to the (Q2) query, which asks about the professional ice hockey establishment is 1917 (present in the false negative). However, the ground truth incorrectly states 1997.

\paragraph{3. The query may be too generic or ambiguous.} In a substantial amount of training pairs in \msmarco, we find that the training query is rather ambiguous, leading to many false negatives being detected. E.g., for the (Q3) query, all passages--including both the ground truth and false negatives--are relevant, as they each correctly define ``yin and yang'' but with different interpretations.

\paragraph{4. False negatives can be partially correct.} Not all detected false negatives are entirely non-relevant to the query. E.g., one false negative is partially relevant to (Q4), which asks about organizations associated with Charles, the Prince of Wales.

\section{Conclusion}

In this work, we emphasize the importance of clean training datasets.
First, we showed that certain datasets can negatively impact model effectiveness when fine-tuned across a huge collection with many training pairs. Dataset pruning removes 57.5\% (8 datasets out of 15) and improves the model accuracy on \beir by even 1.0 point and making the dataset 2.35$\times$ smaller. 
Next, after pruning, we observed the issue of false hard negatives in the remaining training datasets, where passages in the hard negative list are misclassified and are relevant to the query. We presented RLHN, an effective cascading LLM approach for relabeling hard negatives as ground truth or positives. 

Using RLHN, both retrievers and rerankers consistently improved their model generalization on \beir and zero-shot \ab evaluations, as supported by human annotation results. RLHN outperforms hard-negative sampling baseline and is comparable to cross-encoder distillation.

\section*{Limitations}
Even though we propose an effective technique to identify and relabel false hard negatives with RLHN, no technique is perfect and has its limitations. Making those explicit is a critical point in understanding the RLHN results and improvements, and for future work, to propose even better detection techniques.

\paragraph{1. False positives in training datasets.} Detecting and relabeling false positives in training datasets is an important avenue of potential research. However, we avoid checking for false positives, as these labels are trustworthy, provided by a human assessor, who can have a different preference than the LLM itself. False positives might occur in a dataset due to human errors in existing datasets, but we suspect both the importance and frequency of detected false positives to be much lower than false negatives. 

\paragraph{2. Cleaning extremely large training datasets.} The maximum training dataset size that we covered in our work contained $\leq$1M training pairs. This is a reasonable dataset size to apply RLHN within a strict compute budget. Cleaning extremely large training datasets (for example, containing between ~1--10M training pairs) is not feasible, as it may require a very high computation budget, with detection using GPT-4o. In the future, we wish to experiment with open-source LLMs, such as Qwen-3~\cite{yang2025qwen3technicalreport}, as an alternative in our LLM cascading pipeline, allowing relabeling of extremely large training datasets.

\paragraph{3. Multilingual and long-context document retrieval datasets.} A majority of the training datasets included in the BGE training collection have average document lengths up to a few hundred words, roughly equivalent to a few paragraphs. Applying RLHN to clean long-context document retrieval datasets, such as MLDR~\cite{chen:2024} and multilingual training datasets, such as MIRACL~\cite{zhang-etal-2023-miracl}, would be highly relevant in the future.

\paragraph{4. Multi-vector retrieval models.} A popular suite of retrieval models includes multi-vector models, such as ColBERT~\cite{colbert, santhanam:2022}, representing queries and documents by multiple contextualized token-level embeddings. In our work, we limited our experiments to dense retrievers and rerankers. We keep RLHN with an extension to multi-vector models as future work, using a training repository such as PyLate~\cite{PyLate}. 

\section*{Acknowledgments}
This research was supported in part by the Natural Sciences and Engineering Research Council (NSERC) of Canada.
Additional funding is provided by Microsoft via the Accelerating Foundation Models Research program.

\bibliography{camera_ready}

\appendix

\clearpage

\section{Pretrained or Backbone Choice}\label{ap:backbone}
We experimented with several pretrained or base model choices. In particular, we focused on fine-tuning recently introduced encoder models such as ModernBERT~\cite{warner:2024} to decoder-based large language models such as Qwen-2.5 (less than <500M parameters). We fine-tune each backbone on the whole BGE retrieval training subset (15 datasets \& 1.6M training pairs) for up to 3 training epochs with different hyperparameters to fit the training with 4$\times$A6000 GPUs. We plot the model configurations and training settings in \autoref{tab:backbone-choice}.

\smallskip
\noindent\textbf{Validation results.} From \autoref{tab:backbone-choice}, we observe that encoder models pre-trained such as E5-base or E5-large achieve the highest nDCG@10 scores on four \beir datasets. These outperform recent backbones such as ModernBERT-base~\cite{warner:2024} or even smaller-sized LLMs such as Qwen-2.5 (0.5B). 
This anecdotally confirms that the unsupervised pre-training stage in E5 pretrained models is useful and necessary for achieving a competitive nDCG@10 score on \beir. 
Since fine-tuning E5 (large) is around 2$\times$ slower than fine-tuning E5 (base), we run our main experiments on E5 (base) due to computational budget constraints.

\begin{table}[ht]
\centering
\footnotesize
\resizebox{0.48\textwidth}{!}{
\begin{tabular}{lrrrr}
\toprule
\textbf{Dataset} & $\sim$\textbf{100K} & $\sim$\textbf{250K} & $\sim$\textbf{400K} & $\sim$\textbf{680K} \\
\midrule
\msmarco  & 49,571 & 145,000 & 210,000 & 485823 \\
\hotpotqa & 10,250 & 30,000 & 84,516 & 84516 \\
\nq &  6110 & 30,000 & 58,568 & 58,568 \\
\fever & 8017 & 28,755 & 28,755 & 28,755 \\
\scidocsrr  & 12,654 & 12,654 & 12,654 & 12,654 \\
\fiqa & 5500 & 5,500 & 5,500 & 5,500 \\
\arguana & 4065 & 4,065 & 4,065 & 4,065 \\ \midrule
\textbf{Total Pairs} & \textbf{96,167} & \textbf{255,974} & \textbf{404,058} & \textbf{679,881} \\
\bottomrule
\end{tabular}}
\caption{Training pair distribution across seven datasets for four configurations: 100K, 250K, 400K, and 680K.}
\label{tab:label-distribution}
\end{table}
\begin{table*}[t]
\centering   
\small
\resizebox{\textwidth}{!}{%
\begin{tabular}{lcccccccc|rrrr}
\toprule
\textbf{Backbone} & \textbf{\#Params} & \textbf{\#Layers} & \textbf{Hidden Size} & \textbf{Pool} & \textbf{LR} & \textbf{Batch Size} & \textbf{Epoch} & \textbf{Time Taken} & \textbf{COVID} & \textbf{NFC.} & \textbf{FiQA} & \textbf{SciFact} \\ 
\midrule
E5-large (unsup.) \cite{wang:2022} & 330M & 24 & 1024 & \emph{mean} & $1e$ -- 5 & 128 x 8 x 4 & 3 & $\sim$ 36 hours & \underline{0.712} & \textbf{0.383} & \textbf{0.475} & \textbf{0.747} \\
ModernBERT-base \cite{warner:2024} & 149M & 22 & 768 & \emph{mean} & $2e$ -- 4 & 256 x 8 x 4 & 3 & $\sim$ 12 hours & 0.560 & 0.279 & 0.440 & 0.602 \\ 
E5-base  (unsup.) \cite{wang:2022} & 110M & 12 & 768 & \emph{mean} & $2e$ -- 5 & 256 x 8 x 4 & 3 & $\sim$ 18 hours & \textbf{0.731} & \underline{0.381} & \underline{0.444} & \underline{0.728} \\
E5-small (unsup.) \cite{wang:2022} & 33M & 12  & 384 & \emph{mean} & $3e$ -- 5 & 256 x 8 x 4 & 3 & $\sim$ 13 hours & 0.667 & 0.349 & 0.420 & 0.698 \\  \midrule
Qwen-2.5-0.5B~\cite{qwen:2024} & 500M & 24 & 896 & \emph{last} & $1e$ -- 5 & 96 x 8 x 4 & 3 & $\sim$ 36 hours & 0.503 & 0.356 & 0.417 & 0.692 \\
SmolLM2-360M~\cite{ben-allal:2025} & 360M & 32 & 960 & \emph{last} & $1e$ -- 5 & 96 x 8 x 4 & 3 & $\sim$ 33 hours & 0.670 & 0.336 & 0.355 & 0.635 \\
SmolLM2-135M~\cite{ben-allal:2025} & 135M & 30 & 576 & \emph{last} & $1e$ -- 5 & 128 x 8 x 4 & 3 & $\sim$ 24 hours & 0.668 & 0.327 & 0.304 & 0.608 \\
\bottomrule
\end{tabular}}
\caption{
Model configuration, training settings, and retrieval results (nDCG@10) for backbone models fine-tuned on the BGE-training dataset (1.6M training pairs) and evaluated on four datasets from the BEIR benchmark. The models are sorted according to parameter size; The best score is highlighted as \textbf{bold}, the second best is \underline{underlined}. COVID denotes the TREC-COVID dataset and NFC. denotes the NFCorpus dataset.}
\label{tab:backbone-choice}
\end{table*}

\section{Leave-One-Dataset-Out Results} \label{ap:leaving-one-dataset-out}
We provide detailed scores for leave-one-dataset-out (\autoref{fig:beir-leave-one-out-scores}) in \autoref{tab:leave-one-out},
where we fine-tune E5-base retriever models on:\
\begin{enumerate}[leftmargin=30pt,align=left]
    \item[\textbf{Part~(a):}] no datasets;
    \item[\textbf{Part~(b):}] all 15 datasets;
    \item[\textbf{Part~(c):}] all 15 datasets but one left-out dataset;
    \item[\textbf{Part~(d):}] 7 datasets with the most significant effectiveness drop after being removed;
\end{enumerate}

\begin{figure*}[htb]
\begin{mdframed}[font=\footnotesize, roundcorner=10pt, linecolor=purple, linewidth=1pt, innerleftmargin=10pt, innerrightmargin=10pt, innertopmargin=5pt, innerbottommargin=5pt]
\textbf{SYSTEM:} Given (1) a search question, (2) a relevant ground-truth document, (3) and a set of unrelated documents that may appear in any system's response to that question.
Your task is to evaluate whether any of the unrelated documents are relevant compared to the ground-truth document in answering the question.
A document is only considered \textbf{relevant} to the question if it provides sufficient information in answering the question. \\

\#\# \emph{Input} \\

You will receive: \\
\mytab 1.~\emph{question}: The question that the to-be-judged documents will be evaluated on. \\
\mytab 2.~\emph{ground\_truth}: A pre-validated document judged as \textbf{most relevant} to the question. This document can answer the question and should be used as a guide for your analysis. \\
\mytab 3.~\emph{documents}: A set of unrelated documents which may not be relevant in answering the question. \\

You will first read the question and carefully analyze each unrelated documents provided to you. \\
Read every question and unrelated document carefully as you would when proofreading. \\

\#\# \emph{Criteria} \\

Use the following criteria to judge the relevance of each document: 

\mytab - \emph{Relevant}: A document is considered \textbf{relevant} to the question if it provides sufficient information in answering the question, containing \textbf{all} necessary parts highlighted in the ground truth. \\
\mytab - \emph{Not Relevant}: The document does not answer the question and \textbf{does not} provide information in entailing parts present in the ground truth. \\

\#\# \emph{Output} \\

Follow these detailed steps and output your reasoning for each step wrapped for each respective XML tag below: \\
\mytab 1.~You should think and provide your reasoning under \texttt{<thinking>} [ \ldots ] \texttt{</thinking>} on \textbf{why} and \textbf{how} if an unrelated document is \textbf{relevant} following the criteria above. \\
\mytab 2. Next, for all unrelated documents which are found to be \textbf{relevant}, compare them against the ground truth (\texttt{<ground\_truth>}) document in answering the question under \texttt{<preference>} [ \ldots ] \texttt{</preference>} tokens. \\
\mytab 3. Finally, output the list of documents which are (1) relevant and (2) prefer better or equal under the XML tag (\texttt{<better>}) or worse (\texttt{<worse>}) than the ground truth (\texttt{<ground\_truth>}) document for answering the question in \texttt{<verdict>} [ \ldots ] \texttt{</verdict>}. Output [ ] if none of the documents are found to be relevant. \\

Follow strictly the format below: \\
\texttt{<thinking>}
Evaluate the reasoning individually for all unrelated documents to answer the question \\
\mytab Doc (1): output the reasoning here \\
\mytab Doc (2): output the reasoning here \\
\mytab \dots \\
\texttt{</thinking>} \\
\texttt{<preference>} Compare the ground truth and every \textbf{relevant} document individually to answer the question \\
\mytab Doc (1): compare the relevance of Doc (1) with the \texttt{<ground\_truth>} document here, which is more preferred? \\
\mytab \dots \\
\texttt{</preference>} \\
\texttt{<verdict>} \\
\mytab \texttt{<better>} Preferred over or equally as ground truth: [Doc (2) \ldots] \texttt{</better>}, \\
\mytab \texttt{<worse>} Relevant but not preferred over ground truth:  [Doc (1) \ldots] \texttt{</worse>} \\
\texttt{</verdict>} \\

----------- \\
\texttt{<question>}
\{$question$\}
\texttt{</question>} \\
\texttt{<ground\_truth>}
\{$ground\_truth$\}
\texttt{</ground\_truth>} \\
\texttt{<documents>}
\{$documents$\}
\texttt{</documents>}
\end{mdframed}
\caption{Prompt used in RLHN with \gptmini and \gpt for relabeling hard negatives for all BGE training datasets. Certain texts above in the prompt are bolded and tab-aligned to assist with reading. For both \gptmini (stage 1) and \gpt (stage 2) experiments, we consider negatives present within the \texttt{<better>} and \texttt{</better>} tags as false negatives. However, a training instance with any hard negative in either \texttt{<better>} and \texttt{</better>} or \texttt{<worse>} and \texttt{</worse>} tags in the first stage output (\gptmini judge) was forwarded to the second stage (\gpt judge) in the RLHN framework.}
\label{fig:prompt}
\end{figure*}

\begin{table*}[!ht]
\centering
\resizebox{\textwidth}{!}{
\begin{tabular}{llc|cccccccccccccc|cccc}
\toprule
& \textbf{Setting} & \rotatebox[origin=l]{90}{\textbf{Training Pairs}} & \rotatebox[origin=l]{90}{\textbf{TREC-COVID}} & \rotatebox[origin=l]{90}{\textbf{NFCorpus}} & \rotatebox[origin=l]{90}{\textbf{NQ}} & \rotatebox[origin=l]{90}{\textbf{HotpotQA}} & \rotatebox[origin=l]{90}{\textbf{FiQA-2018}} & \rotatebox[origin=l]{90}{\textbf{ArguAna}} & \rotatebox[origin=l]{90}{\textbf{Touch\'e-2020}} & \rotatebox[origin=l]{90}{\textbf{DBPedia}} & \rotatebox[origin=l]{90}{\textbf{SCIDOCS}} & \rotatebox[origin=l]{90}{\textbf{FEVER}} & \rotatebox[origin=l]{90}{\textbf{Climate-FEVER}} & \rotatebox[origin=l]{90}{\textbf{SciFact}} & \rotatebox[origin=l]{90}{\textbf{TREC-NEWS}} & \rotatebox[origin=l]{90}{\textbf{Robust04}} & \rotatebox[origin=l]{90}{\textbf{Avg. 14}} & \rotatebox[origin=l]{90}{\hlgreen{\textbf{Improved}}} & \rotatebox[origin=l]{90}{\hlred{\textbf{Reduced}}} & \rotatebox[origin=l]{90}{\textbf{Keep Dataset?}} \\
\midrule
\textbf{(a)} & \textbf{Pre-trained (Only)} & 0 & 0.610 & 0.358 & 0.390 & 0.524 & 0.401 & 0.422 & 0.169 & 0.354 & 0.211 & 0.634 & 0.154 & 0.737 & 0.441 & 0.416 & 0.416 & - & - & - \\ 
\textbf{(b)} & \textbf{(ALL) Training Pairs} & 1.60M & 0.731 & 0.381 & 0.595 & 0.726 & 0.444 & 0.652 & 0.181 & 0.437 & 0.233 & 0.871 & 0.370 & 0.728 & 0.434 & 0.477 & 0.519 & - & -  & - \\ \midrule
& \textbf{w/o ELI5} & 1.27M & \hlgreen{0.772} & \hlred{0.378} & \hlred{0.593} & \hlgreen{0.728} & \hlred{0.424} & \hlorange{0.652} & \hlgreen{0.213} & \hlred{0.434} & \hlgreen{0.235} & \hlred{0.868} & \hlgreen{0.377} & \hlgreen{0.734} & \hlgreen{0.469} & \hlorange{0.478} & \hlgreen{0.525} & 7 & 5 & \xmark \\
& \textbf{w/o FEVER} & 1.57M & \hlgreen{0.748} & \hlorange{0.379} & \hlgreen{0.598} & \hlorange{0.725} & \hlgreen{0.446} & \hlred{0.647} & \hlred{0.175} & \hlred{0.434} & \hlorange{0.234} & \hlred{0.787} & \hlred{0.240} & \hlgreen{0.749} & \hlred{0.423} & \hlgreen{0.483} & \hlred{0.505} & 6 & 5 & \cmark \\
& \textbf{w/o HotpotQA} & 1.51M & \hlred{0.724} & \hlorange{0.381} & \hlgreen{0.600} & \hlred{0.642} & \hlgreen{0.449} & \hlorange{0.652} & \hlred{0.178} & \hlred{0.425} & \hlorange{0.232} & \hlred{0.863} & \hlred{0.358} & \hlred{0.725} & \hlgreen{0.441} & \hlgreen{0.489} & \hlred{0.511} & 4 & 7 & \cmark \\
& \textbf{w/o \msmarco Document} & 1.23M & \hlgreen{0.742} & \hlorange{0.380} & \hlred{0.586} & \hlorange{0.726} & \hlorange{0.445} & \hlgreen{0.656} & \hlred{0.175} & \hlred{0.435} & \hlgreen{0.235} & \hlred{0.866} & \hlred{0.347} & \hlgreen{0.742} & \hlgreen{0.458} & \hlgreen{0.490} & \hlorange{0.520} & 6 & 5 & \xmark \\
& \textbf{w/o Stack Overflow (Dup.)} & 1.58M & \hlred{0.720} & \hlred{0.379} & \hlred{0.593} & \hlorange{0.726} & \hlorange{0.444} & \hlred{0.650} & \hlred{0.174} & \hlorange{0.436} & \hlgreen{0.235} & \hlorange{0.870} & \hlred{0.368} & \hlorange{0.729} & \hlred{0.431} & \hlgreen{0.487} & \hlred{0.517} & 7 & 2 & \xmark \\
& \textbf{w/o Trivia QA} & 1.54M & \hlred{0.729} & \hlorange{0.380} & \hlorange{0.595} & \hlgreen{0.730} & \hlgreen{0.450} & \hlred{0.647} & \hlred{0.174} & \hlgreen{0.440} & \hlorange{0.234} & \hlorange{0.870} & \hlgreen{0.382} & \hlgreen{0.731} & \hlgreen{0.443} & \hlgreen{0.481} & \hlorange{0.520} & 7 & 3 & \xmark \\
& \textbf{w/o NLI} & 1.60M & \hlred{0.729} & \hlorange{0.380} & \hlorange{0.594} & \hlorange{0.726} & \hlorange{0.445} & \hlorange{0.652} & \hlred{0.177} & \hlorange{0.437} & \hlorange{0.233} & \hlorange{0.870} & \hlred{0.368} & \hlorange{0.728} & \hlgreen{0.436} & \hlorange{0.477} & \hlorange{0.518} & 1 & 3 & \xmark \\
\textbf{(c)} & \textbf{w/o SQuAD} & 1.51M & \hlred{0.709} & \hlred{0.379} & \hlgreen{0.598} & \hlred{0.723} & \hlorange{0.445} & \hlgreen{0.654} & \hlorange{0.181} & \hlorange{0.437} & \hlorange{0.234} & \hlorange{0.872} & \hlgreen{0.376} & \hlorange{0.729} & \hlgreen{0.439} & \hlgreen{0.481} & \hlorange{0.518} & 5 & 3 & \xmark \\
& \textbf{w/o ArguAna} & 1.59M & \hlgreen{0.736} & \hlorange{0.381} & \hlgreen{0.598} & \hlgreen{0.728} & \hlgreen{0.448} & \hlred{0.434} & \hlred{0.174} & \hlred{0.434} & \hlorange{0.234} & \hlorange{0.871} & \hlgreen{0.378} & \hlgreen{0.731} & \hlgreen{0.445} & \hlgreen{0.486} & \hlred{0.506} & 8 & 3 & \cmark \\
& \textbf{w/o FIQA-2018} & 1.59M & \hlred{0.728} & \hlorange{0.380} & \hlorange{0.596} & \hlorange{0.727} & \hlred{0.428} & \hlgreen{0.658} & \hlred{0.174} & \hlorange{0.436} & \hlgreen{0.235} & \hlorange{0.871} & \hlorange{0.370} & \hlorange{0.729} & \hlorange{0.433} & \hlorange{0.477} & \hlred{0.517} & 3 & 2 & \cmark \\
& \textbf{w/o \msmarco Passage} & 1.11M & \hlred{0.699} & \hlred{0.377} & \hlred{0.551} & \hlgreen{0.730} & \hlred{0.440} & \hlred{0.650} & \hlred{0.162} & \hlred{0.407} & \hlgreen{0.237} & \hlred{0.869} & \hlred{0.338} & \hlgreen{0.733} & \hlred{0.431} & \hlgreen{0.484} & \hlred{0.508} & 3 & 10 & \cmark \\
& \textbf{w/o NQ} & 1.54M & \hlgreen{0.745} & \hlorange{0.381} & \hlred{0.553} & \hlgreen{0.728} & \hlgreen{0.451} & \hlgreen{0.659} & \hlred{0.178} & \hlred{0.435} & \hlorange{0.234} & \hlred{0.867} & \hlorange{0.369} & \hlorange{0.728} & \hlorange{0.435} & \hlred{0.472} & \hlred{0.517} & 5 & 4 & \cmark \\
& \textbf{w/o Quora} & 1.54M & \hlgreen{0.759} & \hlorange{0.382} & \hlgreen{0.599} & \hlorange{0.727} & \hlgreen{0.451} & \hlorange{0.653} & \hlgreen{0.185} & \hlorange{0.436} & \hlorange{0.234} & \hlred{0.867} & \hlorange{0.371} & \hlorange{0.729} & \hlgreen{0.436} & \hlgreen{0.481} & \hlgreen{0.522} & 6 & 1 & \xmark \\
& \textbf{w/o \scidocsrr} & 1.59M & \hlgreen{0.733} & \hlred{0.378} & \hlorange{0.595} & \hlorange{0.727} & \hlgreen{0.447} & \hlgreen{0.662} & \hlred{0.178} & \hlorange{0.436} & \hlred{0.201} & \hlred{0.868} & \hlgreen{0.374} & \hlgreen{0.740} & \hlorange{0.434} & \hlred{0.475} & \hlorange{0.518} & 5 & 4 & \cmark \\
& \textbf{w/o STS} & 1.60M & \hlred{0.718} & \hlred{0.379} & \hlorange{0.596} & \hlorange{0.727} & \hlgreen{0.446} & \hlorange{0.652} & \hlred{0.177} & \hlorange{0.437} & \hlorange{0.234} & \hlred{0.867} & \hlorange{0.369} & \hlorange{0.729} & \hlorange{0.435} & \hlorange{0.478} & \hlred{0.517} & 1 & 4 & \xmark \\ \midrule
\textbf{(d)} & \textbf{7 Datasets Pruned (\cmark)} & 680K & \hlgreen{0.781} & \hlred{0.376} & \hlred{0.593} & \hlgreen{0.728} & \hlred{0.421} & \hlgreen{0.664} & \hlgreen{0.242} & \hlgreen{0.440} & \hlred{0.204} & \hlgreen{0.875} & \hlgreen{0.397} & \hlgreen{0.748} & \hlgreen{0.467} & \hlred{0.464} & \hlgreen{0.529} & 9 & 5 & - \\ 
\bottomrule
\end{tabular}}
\caption{Retrieval results measuring nDCG@10 on 14 datasets in the BEIR benchmark by fine-tuning E5 (base) by \textbf{leaving out one training dataset at a time} and fine-tuning the rest.  \hlgreen{Improved} denotes E5 (base) with a nDCG@10 better than $+1$ point, \hlred{Reduced} with a nDCG@10 worse than $-1$ point, and \hlorange{No Change} within the $\pm1$ point range, compared to part (b) E5 (base) fine-tuned on ALL Training Pairs. 
Each row in part (c) is fine-tuned on all but one left-out dataset.
Part (c) is fine-tuned on the 7 selected datasets.
}
\label{tab:leave-one-out}
\end{table*}

\begin{figure*}[t]
\centering
\begin{center}
    \includegraphics[trim=0 0 0 0,clip,width=\textwidth]{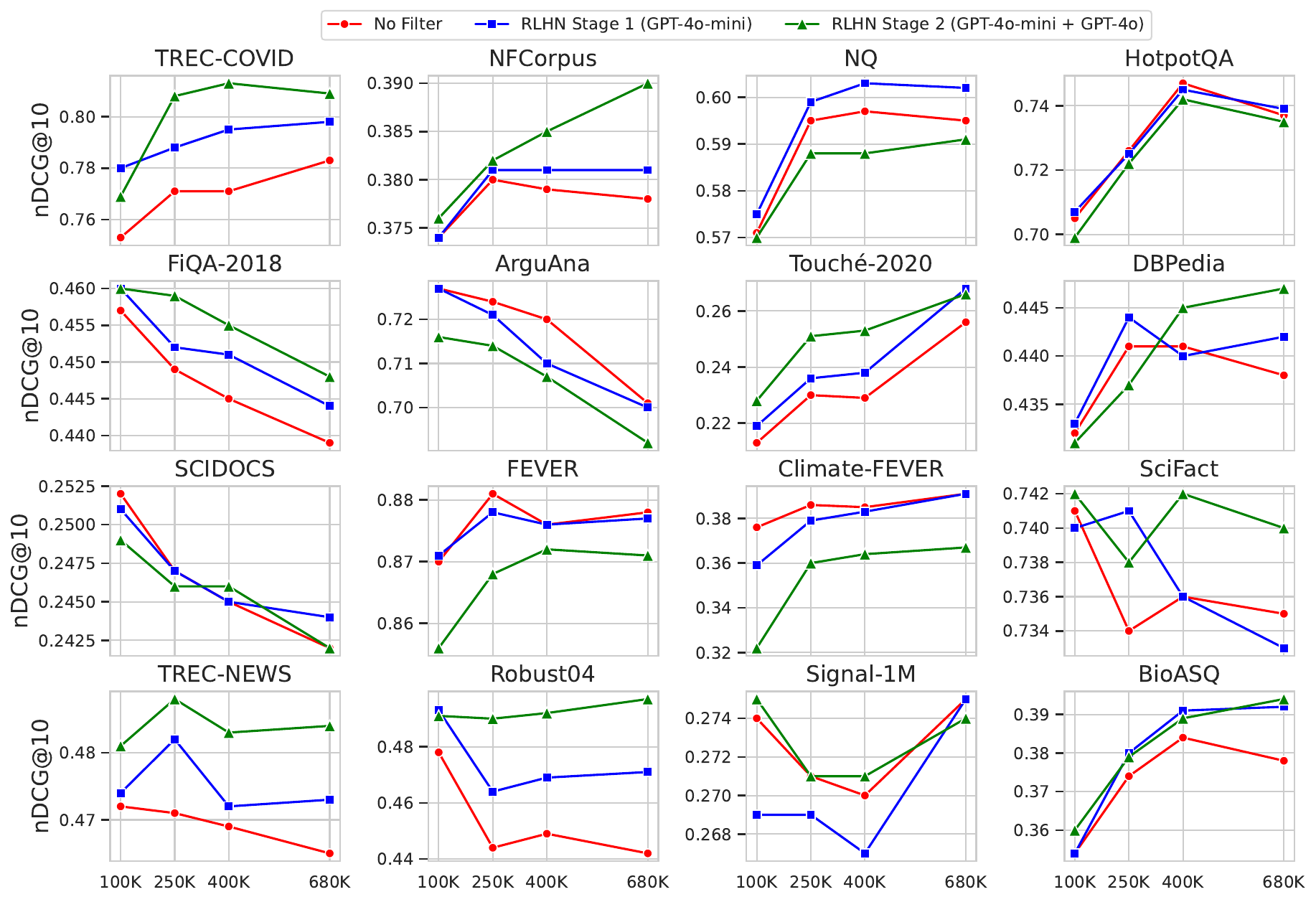}
    \caption{nDCG@10 scores on all 16 \beir datasets by fine-tuning E5 (base) retrieval model on a subset of the 100K, 250K, 400K, and 680K training pairs from both stages 1 and 2, sampled from seven datasets in the BGE collection (listed in \autoref{tab:train_stats}) using the RLHN framework. The training pair distribution is shown in \autoref{tab:label-distribution}.}
    \label{fig:all-scores-beir}
\end{center}
\vspace*{-\baselineskip}
\end{figure*}

\begin{figure*}[t]
\centering
\begin{center}
    \includegraphics[trim=0 0 0 0,clip,width=\textwidth]{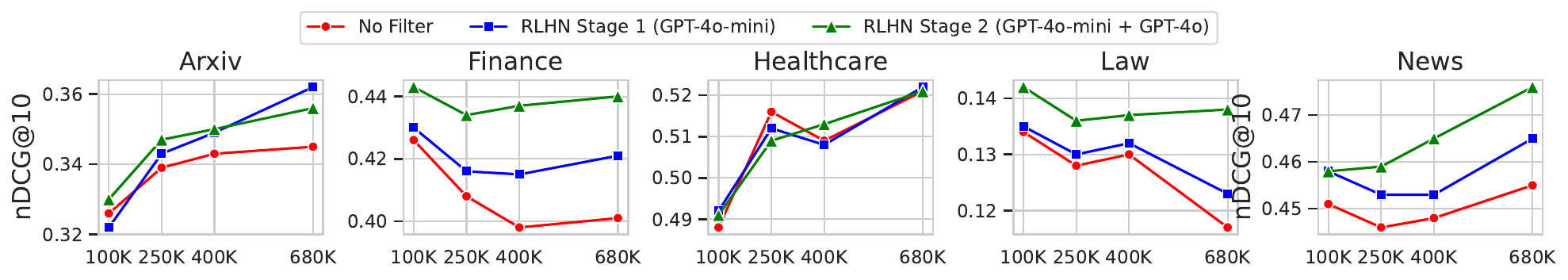}
    \caption{nDCG@10 scores on all 5 \ab datasets by fine-tuning E5 (base) retrieval model on a subset of the 100K, 250K, 400K, and 680K training pairs from both stages 1 and 2, sampled from seven datasets in the BGE collection (listed in \autoref{tab:train_stats}) using the RLHN framework. The training pair distribution is shown in \autoref{tab:label-distribution}.}
    \label{fig:all-scores-air-bench}
\end{center}
\vspace*{-\baselineskip}
\end{figure*}

\begin{figure*}[t]
\centering
\begin{center}
    \includegraphics[trim=0 0 0 0,clip,width=\textwidth]{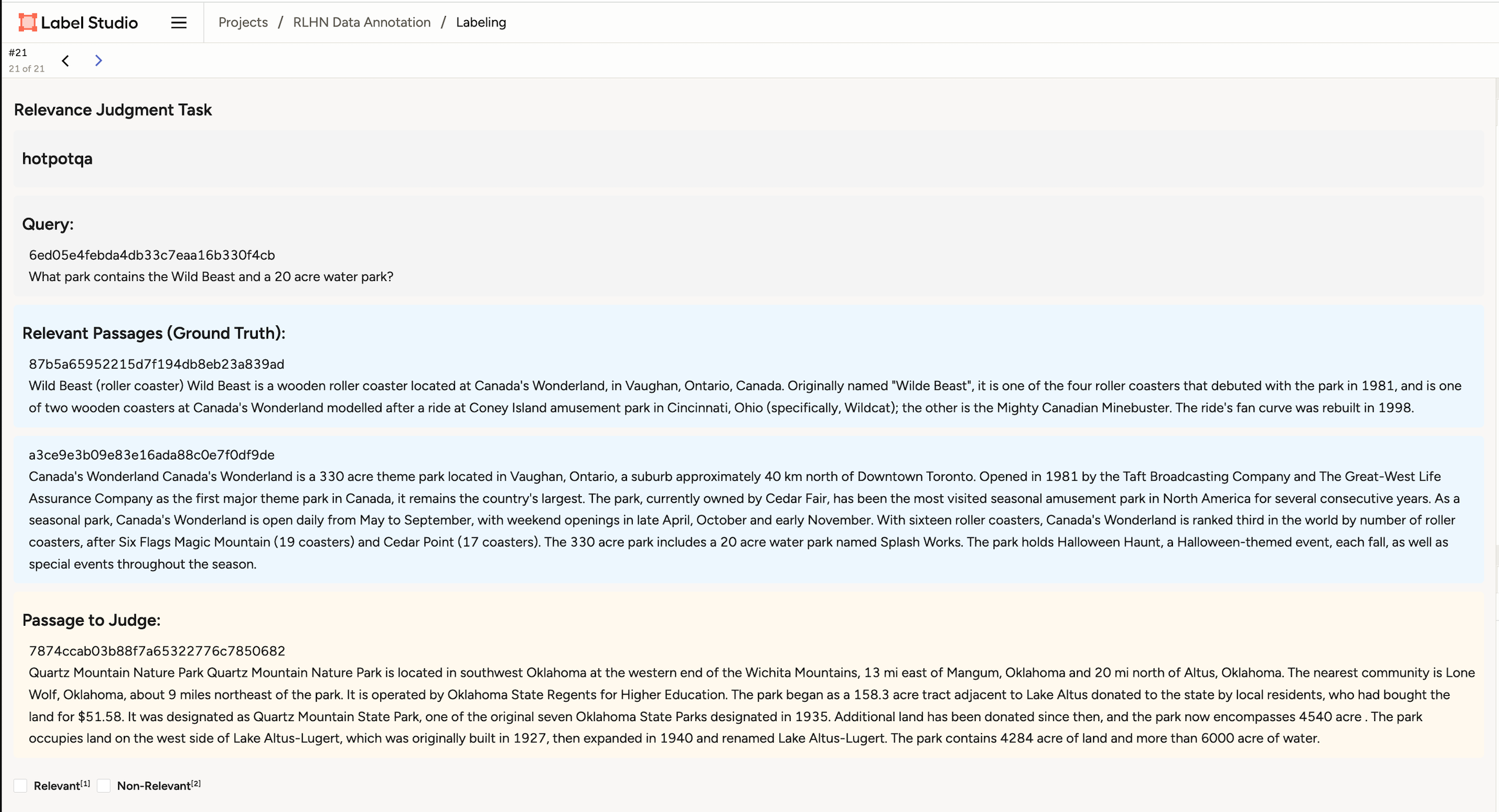}
    \caption{A screenshot of the human validation study conducted via Label Studio. First, the human assessor reads the query (highlighted in grey) and the relevant passages (highlighted in blue). Next, the assessor reads a sequence of hard negative passages one by one (highlighted in yellow) and evaluates the relevancy with the question, marking their decision in the checkbox as either (1) \emph{relevant} or (2) \emph{non-relevant}.}
    \label{fig:label-studio}
\end{center}
\vspace*{-\baselineskip}
\end{figure*}

\begin{figure*}[t]
\centering
\begin{center}
    \includegraphics[trim=0 0 0 0,clip,width=0.8\textwidth]{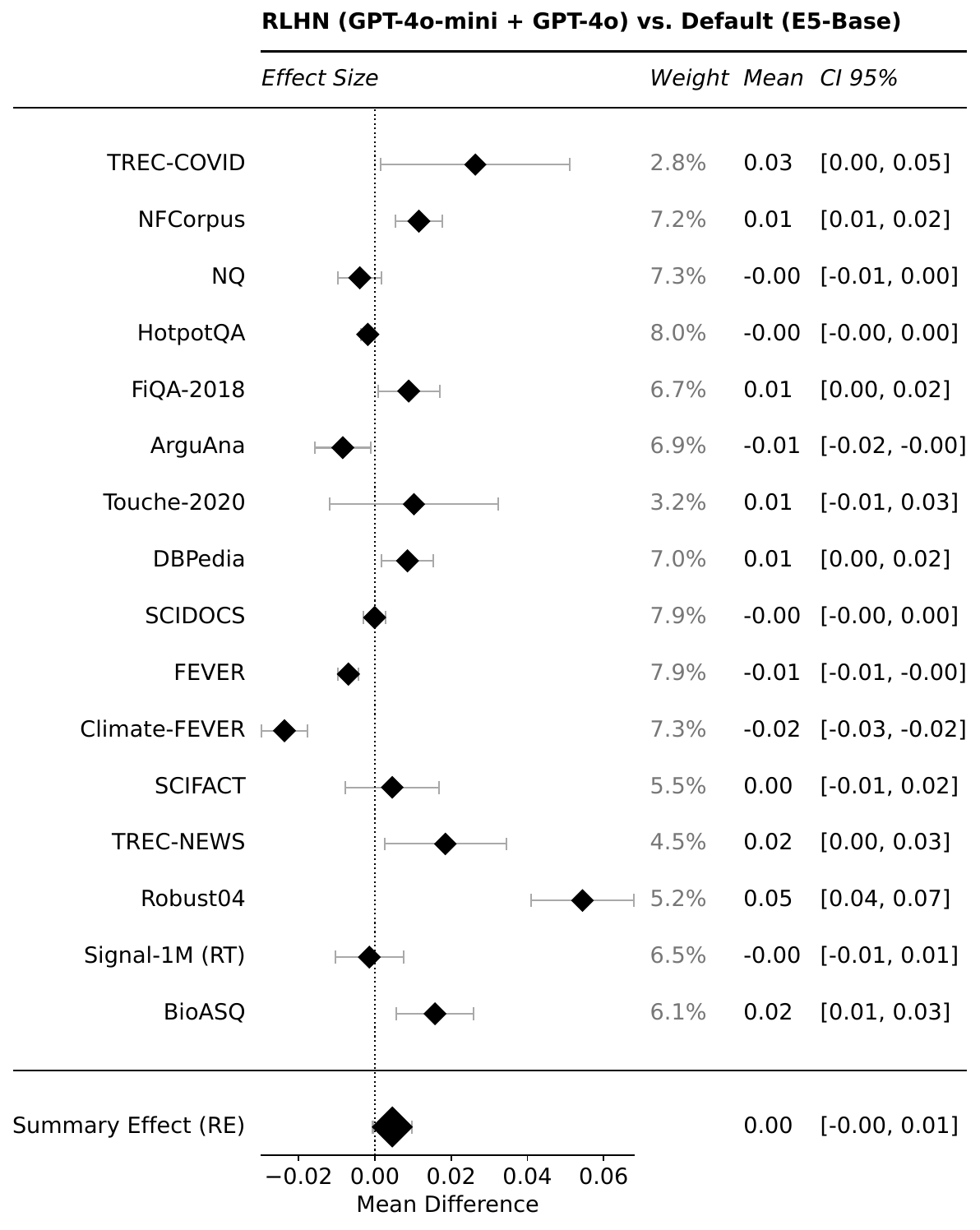}
    \caption{Ranger plot~\cite{sertkan-etal-2023-ranger} showing the statistical significance of improvements observed in RLHN (\gptmini + \gpt) versus the Default setting for the E5 (base) fine-tuned retriever.}
    \label{fig:ranger-plot-e5-base}
\end{center}
\vspace*{-\baselineskip}
\end{figure*}

\begin{figure*}[t]
\centering
\begin{center}
    \includegraphics[trim=0 0 0 0,clip,width=0.8\textwidth]{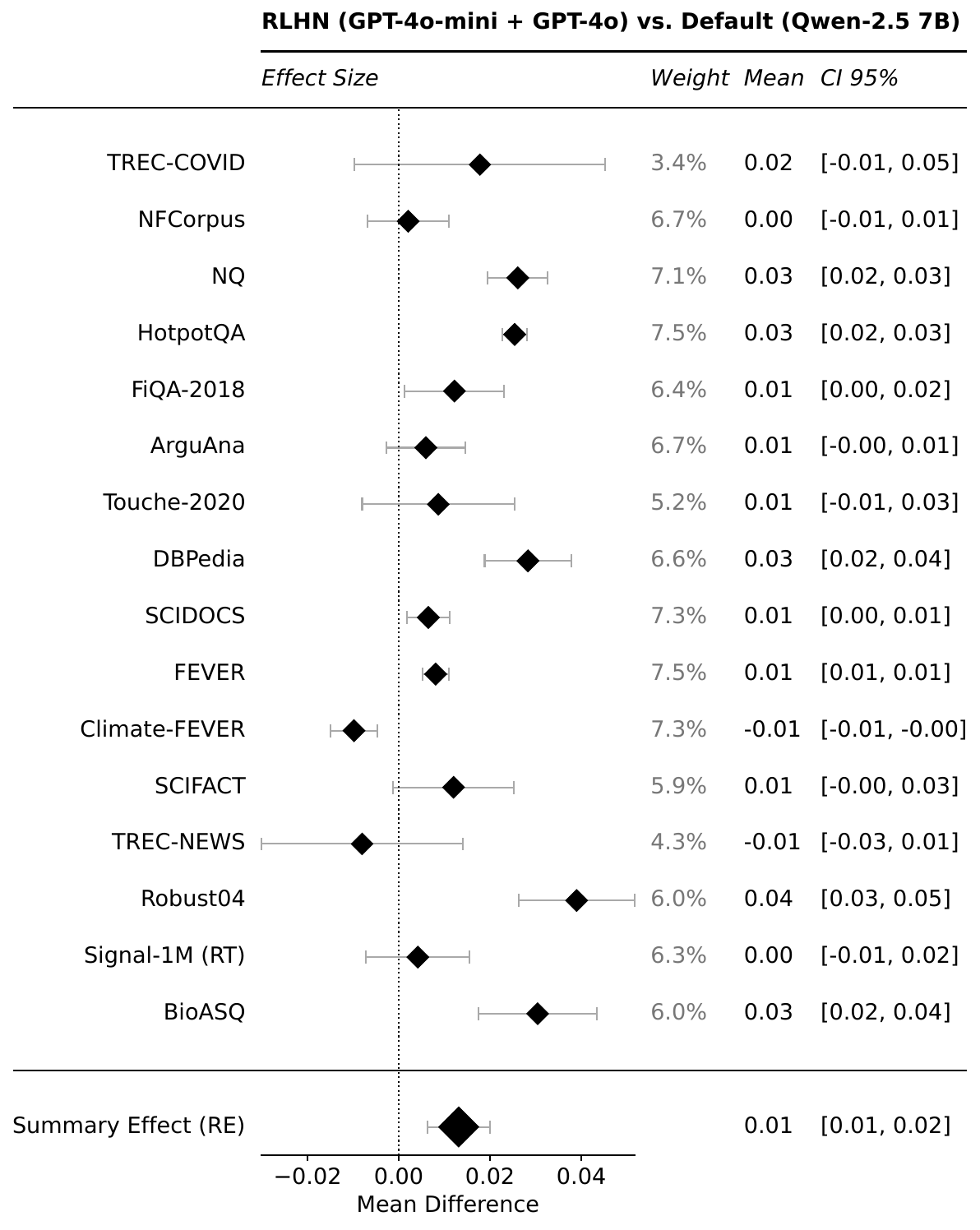}
    \caption{Ranger plot~\cite{sertkan-etal-2023-ranger} showing the statistical significance of improvements observed in RLHN (\gptmini + \gpt) versus the Default setting for the Qwen2.5-7B fine-tuned retriever.}
    \label{fig:ranger-plot-qwen-base}
\end{center}
\vspace*{-\baselineskip}
\end{figure*}

\end{document}